\documentclass[aps,amsmath,prl,twocolumn,superscriptaddress]{revtex4-1}
\usepackage{amsmath,amssymb,graphicx,bm,color,booktabs}
\usepackage{xr}
\usepackage{subfigure}
\usepackage{graphicx}
\usepackage{rotating} 
\usepackage{color}
\usepackage{rotating}
\usepackage{epstopdf}
\usepackage{placeins}
\usepackage{bm}
\extrafloats{100}

\vspace*{0.2in}

\usepackage{scrextend}

\newcommand{\bave}[1]{\left\langle #1 \right\rangle}
\newcommand{\pave}[1]{\left( #1 \right)}

\newcommand{\bea}[1]{ \begin{equation}\begin{aligned} #1 \end{aligned} \end{equation} }

\newcommand{\beaw}[1]{ \begin{widetext} \begin{equation}\begin{aligned} #1 \end{aligned} \end{equation} \end{widetext} }

\newcommand{\F}{\mathcal{F}}
\DeclareMathAlphabet{\mathpzc}{OT1}{pzc}{m}{it}
\newcommand{\E}{\mathcal{E}}

\newcommand{\M}{\mathcal{M}}
\newcommand{\A}{\mathcal{A}}
\newcommand{\I}{\mathcal{I}}
\newcommand{\ra}{\rightarrow}
\usepackage{xargs} 
\usepackage{gensymb} 

\begin{document}
\title{Energetic costs, precision, and efficiency of a biological motor in cargo transport}

\author{Wonseok Hwang}
\affiliation{Korea Institute for Advanced Study, Seoul 02455, Republic of Korea}
\author{Changbong Hyeon}
\email{hyeoncb@kias.re.kr}
\affiliation{Korea Institute for Advanced Study, Seoul 02455, Republic of Korea}


\begin{abstract}
	Molecular motors play pivotal roles in organizing the interior of cells. 
	A motor efficient in cargo transport would move along cytoskeletal filaments with a high speed and a minimal error in transport distance (or time) while consuming a minimal amount of energy. 
	The travel distance of the motor and its variance are, however, physically constrained by the free energy being consumed. 
	A recently formulated thermodynamic principle, called the \emph{thermodynamic uncertainty relation}, offers a theoretical framework for the energy-accuracy trade-off relation ubiquitous in biological processes. According to the relation, a measure $\mathcal{Q}$, the product between the heat dissipated from a motor and the squared relative error in the displacement, has a minimal theoretical bound ($\mathcal{Q} \geq 2 k_B T$), which is approached when the time trajectory of the motor is maximally regular for a given amount of free energy input.
	Here, we use the uncertainty measure ($\mathcal{Q}$) to quantify the transport efficiency of biological motors.
	Analyses on the motility data from several types of molecular motors reveal that $\mathcal{Q}$ is a complex function of ATP concentration and load ($f$).  
	For kinesin-1, $\mathcal{Q}$ approaches the theoretical bound at $f\approx 4$ pN and over a broad range of ATP concentration (1 $\mu$M -- 10 mM), and is locally minimized at [ATP] $\approx$ 200 $\mu$M.
	In stark contrast to the wild type, this local minimum vanishes for a mutant that has a longer neck-linker, and the value of $\mathcal{Q}$ is significantly greater, which underscores the importance of molecular structure. 
	Transport efficiencies of the biological motors studied here are semi-optimized under the cellular condition ([ATP] $\approx 1$ mM, $f=0-1$ pN).
	Our study indicates that among many possible directions of optimization, cytoskeletal motors are designed to operate at a high speed with a minimal error while leveraging their energy resources. 
\end{abstract}

\maketitle      

\section{Introduction}
Biological systems are in nonequilibrium steady states (NESS) in which the energy and material currents flow constantly in and out of the system. 
Subjected to incessant thermal and nonequilibrium fluctuations, 
cellular processes are inherently stochastic and error-prone. 
Biological systems adopt a plethora of error-correcting mechanisms that utilize energy to fix any error deleterious to their functions \cite{sartori2015PRX}.
Trade-off relations between the energetic cost and information processing are ubiquitous in cellular processes, and have been a recurring theme in physics and biology for many decades \cite{Hopfield74PNAS,ehrenberg1980BJ,bennett1982IJTP,AlbertsBook,mehta2012PNAS,Lan2012NaturePhysics,banerjee2017PNAS}.

A recent study by Barato and Seifert \cite{barato2015PRL} has formulated a concise inequality known as  the \emph{thermodynamic uncertainty relation}, which quantifies the trade-off between free energy consumption and precision of an observable from dissipative processes in NESS. 
In their study, the uncertainty measure $\mathcal{Q}$ is defined as the product between the energy consumption (heat dissipation, $Q(t)$) of a driven process in the steady state and the squared relative error of an output observable from the process $X(t)$, $\epsilon^2_X(t)=\langle\delta X^2\rangle/\langle X\rangle^2$.
It has been further conjectured that for an arbitrary chemical network formulated by Markov jump processes $\mathcal{Q}$ cannot be smaller than $2k_BT$,  
\begin{align}
	\mathcal{Q}=Q(t)\times \epsilon_X^2(t)\geq 2k_BT. 
	\label{eqn:inequality1}  
\end{align}
The measure $\mathcal{Q}$ quantifies the uncertainty of a dynamic process. 
The smaller the value of $\mathcal{Q}$, the more regular and predictable is the trajectory generated from the process, improving the precision of the output observable. 
In the presence of large fluctuations inherent to cellular processes, harnessing energy into precise motion is critical for accuracy in cellular computation. 
The uncertainty measure $\mathcal{Q}$ can be used to assess the efficiency of suppressing the uncertainty in dynamical process via energy consumption.
The proof and physical significance of this inequality have been discussed \cite{barato2015PRL,Gingrich2016RPL,pietzonka2016PRE,Pigolotti2017PRL,Hyeon2017PRE,Proesmans:2017}. 
Among others, we have  shown that the minimal bound of $\mathcal{Q}$, $2k_BT$, is attained when heat dissipated from the process is normally distributed, such that $P(Q)\sim e^{-Q^2}$ \cite{Hyeon2017PRE}.

Historically, the efficiency of heat engines has been discussed in terms of the thermodynamic efficiency, the aim of which is to maximize the amount of work extracted from two heat reservoirs with different temperatures \cite{lee2017SciReport}. 
For nonequilibrium machines in general driven by chemical forces that are constantly regulated in the live cell, 
the power production could be a more pertinent quantity to maximize. 
Meanwhile, for transport motors in the cell, the uncertainty measure $\mathcal{Q}$ can be used to assess the transport efficiency of a motor (or motors) \cite{dechant2017Arxiv}. 
Because the displacement $l(t)$ (or travel distance) is a natural output observable of interest in the cargo transport, we set $X=l(t)$, which recasts Eq.\ref{eqn:inequality1} into  
\begin{align}
	\mathcal{Q}=\dot{Q}\frac{2D}{V^2}\geq 2k_BT.
	\label{eqn:inequality2}
\end{align}
$\mathcal{Q}$ is minimized by a motor that transports cargos (i) at a high speed ($V\sim \langle l(t)\rangle/t$), (ii) with a small error ($D\sim \langle\delta l(t)^2\rangle/t$) in the displacement (or punctual delivery to a target site), and (iii) with a small energy consumption ($\dot{Q}$).   
Thus, a molecular motor efficient in the cargo transport is characterized by a small $\mathcal{Q}$ with its minimal bound $2k_BT$.

In the present study, we assess the ``transport efficiency'' of several biological motors (kinesin-1, KIF17 and KIF3AB in kinesin-2 family, myosin-V, dynein, and F$_1$-ATPase) in terms of $\mathcal{Q}$, and study how it changes with varying conditions of load ($f$) and [ATP]. 
Of particular interest is to identify the optimal condition for motor efficiency, if any, that $\mathcal{Q}$ is minimized. 
To evaluate $\mathcal{Q}$, one should know $\dot{Q}$, $D$, and $V$ of the system (see Eq.\ref{eqn:inequality2}), which can be obtained using a suitable kinetic network model that can delineate the dynamical characteristics of the system \cite{Hwang2017JPCL}.   
Our analyses on motors show that $\mathcal{Q}$, demonstrating a complex functional dependence on $f$ and [ATP], is sensitive to a subtle variation in motor structure.  
Transport and rotory motors studied here are semi-optimized in terms of $\mathcal{Q}$ under the cellular condition, which alludes to the role of evolutionary pressure that has shaped the current forms of molecular motors in the cell. 

\begin{figure*}[t]
	\centering
	\includegraphics[scale=0.57]{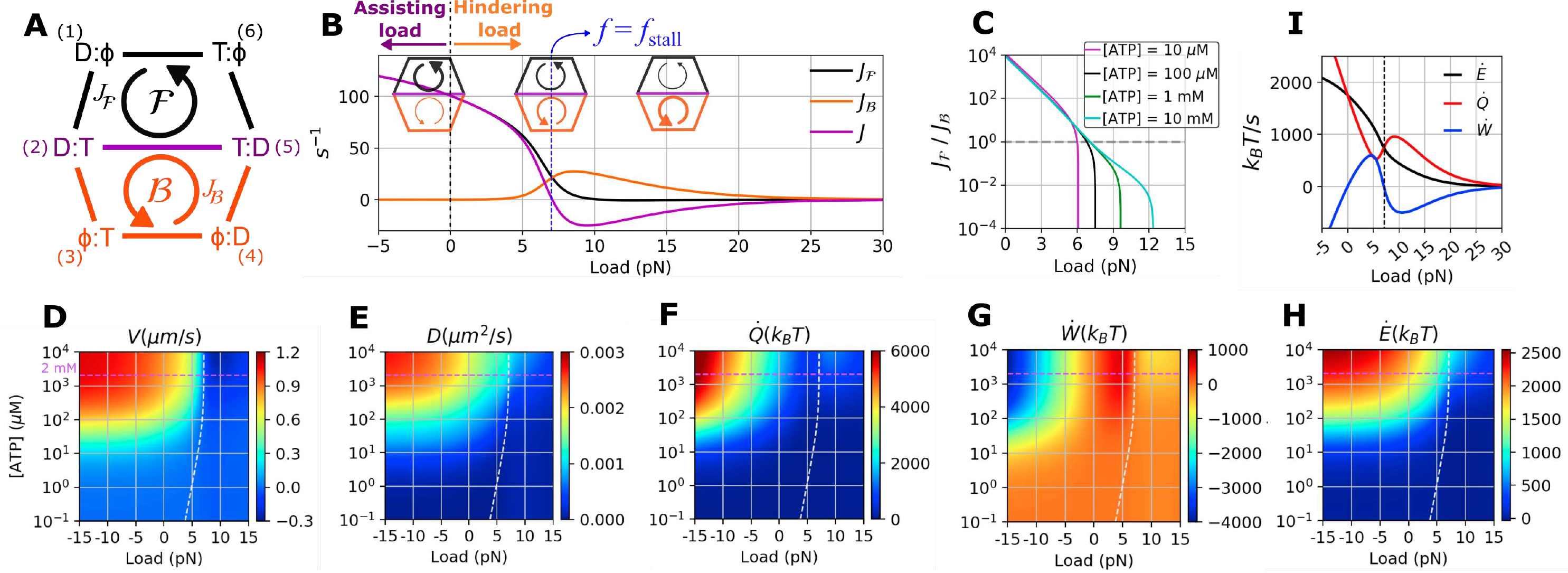}
	\caption{Analysis of kinesin-1 based on the 6-state double-cycle kinetic model. 
		{\bf A.} Schematics of the 6-state double-cycle kinetic network for hand-over-hand dynamics of kinesin-1, where $T$, $D$, and $\phi$ denote ATP-, ADP-bound, and apo state, respectively.
		Through ATP binding [(1)$\rightarrow$(2)], mechanical step [(2)$\rightarrow$(5)], release of ADP [(5)$\rightarrow$(6)], and hydrolysis of ATP [(6)$\rightarrow$(1)], kinesin moves forward in the $\mathcal{F}$-cycle [$(1) \rightarrow (2) \rightarrow (5) \rightarrow (6) \rightarrow (1)$], where it takes a backstep in the $\mathcal{B}$-cycle [$(4) \rightarrow (5) \rightarrow (2) \rightarrow (3) \rightarrow (4)$]. 
		The arrows in the figure depict the direction of reaction currents.
		In both cycles, each chemical step is reversible and the transition rate from the $(i)$-th to $(j)$-th state  is given by $k_{ij}$.
		{\bf B.} Reaction current $J_{\mathcal{F}}$, $J_{\mathcal{B}}$, and $J$ as a function of load. The three cartoons illustrate the amount of current along the $\mathcal{F}$ and $\mathcal{B}$ cycles at each value of load. 
		$f<0$ and $f>0$ correspond to the assisting and hindering load, respectively. 
		{\bf C.} The ratio between the forward and backward fluxes $J_\mathcal{F}/J_\mathcal{B}$ as function of $f$ at fixed [ATP]. 
		The stall forces, determined at $J_\mathcal{F} / J_\mathcal{B}=1$ (dashed line), are narrowly distributed between $f=6-8$ pN.
		({\bf D}--{\bf H}) $V$, $D$, $\dot{Q}$, $\dot{W}$, $\dot{E}$ as a function of $f$ and [ATP]. 
		The white dashed lines demarcate the locus of [ATP]-dependent stall force, and the dashed lines in magenta indicate the condition of [ATP] = 2 mM. 
		{\bf I}. Dependences of $\dot{E}$, $\dot{Q}$, $\dot{W}$ on $f$ at [ATP] = 2 mM. 
	}
	\label{fig_kinesin_6state}
\end{figure*} 

\section{Results}
\subsection{Chemical driving force, steady state current, and heat dissipation of double-cycle kinetic network for kinesin-1}

To study the transport properties of molecular motors, we used experimental data, available in the literature, of $V$ and $D$ under varying conditions of $f$ and [ATP]. 
Once a set of kinetic rate constants $\{k_{ij}\}$ that defines the network model is determined by fitting the data of $V(\{k_{ij}(f,[\text{ATP}])\})$ and $D(\{k_{ij}(f,[\text{ATP}])\})$, it is straightforward to calculate $\dot{Q}(f,[\text{ATP}])$ \cite{Hwang2017JPCL}, and hence $\mathcal{Q}(f,[\text{ATP}])$. 

For the case of kinesin-1, we empoloy
the 6-state kinetic network model \cite{Liepelt07PRL}, consisting of two cycles, $\mathcal{F}$ and $\mathcal{B}$ (Fig. \ref{fig_kinesin_6state}A). 
Although the conventional (N=4)-state unicyclic kinetic model \cite{Fisher1999_PNAS,Fisher01PNAS} confers a similar result with the 6-state double-cycle network model at small $f$ (compare Figs.\ref{fig_kinesin_6state} and 	\ref{fig_kinesin_uni}), 
the unicyclic model is led to a physically problematic interpretation especially when the molecular motor is stalled and starts taking backsteps at large hindering load \cite{Astumian1996BJ,Liepelt07PRL,Hyeon09PCCP}. 
As explicated previously in Ref. \cite{Hyeon09PCCP}, the backstep in the unicyclic network, by construction, is produced by a reversal of the forward cycle, which implies that the backstep is always realized via the synthesis of ATP from ADP and P$_i$.  
More importantly, in calculating $\dot{Q}$ from kinetic network, the unicyclic network results in $\dot{Q}=0$ under the stall condition, which however contradicts the physical reality; an idling car still burns fuel and dissipates heat, thus $\dot{Q}\neq 0$. 
To build a more physically sensible model that considers the possibility of ATP-induced (fuel-burning) backstep, we extend the unicyclic network into a multi-cyclic one which takes into account an ATP-consuming stall, i.e., a futile cycle \cite{Liepelt07PRL,Yildiz08Cell,Hyeon09PCCP,clancy2011NSMB}. 

The proposed double-cycle network scheme is physically more sensible and general than unicyclic schemes in that it can accommodate 4 different possibilities for the kinetic paths: 
(i) ATP-hydrolysis induced forward step; 
(ii) ATP-hydrolysis induced backward step; 
(iii) ATP-synthesis induced forward step;
(iv) ATP-synthesis induced backward step.  
With the kinetic rate constants determined for kinesin-1 using the double-cycle model, 
the kinesin-1 predominantly moves forward through the $\mathcal{F}$-cycle under small hindering ($f>0$) or assisting load ($f<0$), whereas it takes a backstep through the $\mathcal{B}$-cycle under a large hindering load. 
In principle, the steady-state reaction current within the $\mathcal{F}$-cycle, $J_{\mathcal{F}}$, itself is decomposed into the forward ($J^+_{\mathcal{F}}$) and backward current ($J^-_{\mathcal{F}}$), such that $J_{\mathcal{F}}=J^+_{\mathcal{F}}-J^-_{\mathcal{F}}>0$. 
Although a backstep could be realized through an ATP synthesis \cite{Hackney05PNAS}, corresponding to $J_{\mathcal{F}}^-$, a theoretical analysis \cite{Hyeon09PCCP} on experimental data \cite{Nishiyama02NCB,Cross05Nature} suggest that such backstep current (ATP synthesis induced backstep, $J^-_{\mathcal{F}}$) is negligible in comparison with $J^+_{\mathcal{B}}$ (ATP hydrolysis induced backstep).

We illuminate the dynamics realized in the double-cycle network by calculating $J_{\mathcal{F}}$ and $J_{\mathcal{B}}$ with increasing $f$ (see Fig.\ref{fig_kinesin_6state}B). 
Without load ($f=0$), kinesin-1 predominantly moves forward ($J_{\mathcal{F}}\gg J_{\mathcal{B}}$). 
This imbalance diminishes as $f$ is increased. 
At stall conditions, the two reaction currents are balanced ($J_\mathcal{F}  = J_\mathcal{B}$), so that the net current $J$ associated with the mechanical stepping defined between the states (2) and (5) vanishes ($J=J_\mathcal{F}-J_\mathcal{B}=0$), but nonvanishing current due to chemistry still remains along the cycle of $\rightarrow (2)\rightarrow (3)\rightarrow (4)\rightarrow (5)\rightarrow (6)\rightarrow (1)\rightarrow (2)\rightarrow$ (see Fig.\ref{fig_kinesin_6state}A). A further increase of $f$ beyond the stall force renders $J_{\mathcal{F}}< J_{\mathcal{B}}$, augmenting the likelihood of backstep.

For a given set of rate constants, it is straightforward to calculate the rates of heat dissipation ($\dot{Q}$), work production ($\dot{W}$), and total energy supply ($\dot{E}$).
The total heat generated from the kinetic cycle depicted in Fig. \ref{fig_kinesin_6state}A is decomposed into the heat generated from two subcycles, $\dot{\mathcal{Q}}_{\mathcal{F}}$ and $\dot{\mathcal{Q}}_{\mathcal{B}}$, each of which is the product of reaction current and affinity \cite{Liepelt07PRL,barato2015PRL,Seifert2012RPP,Qian2005_BiophyChem,Qian_PRE_2004,Qian2010PRE,Wachtel:2015:PRE}
\begin{equation} \label{eq:dotQ}
	\dot{Q}  = \dot{Q}_\mathcal{F} + \dot{Q}_\mathcal{B}= J_\mathcal{F} \mathcal{A}_\mathcal{F} + J_\mathcal{B} \mathcal{A}_\mathcal{B}.  
\end{equation}
Here, the affinities (driving forces) for the $\mathcal{F}$ and $\mathcal{B}$ cycles are 
\begin{align}
	\mathcal{A}_\mathcal{F} = k_B T \log \left( \frac{ k_{12} k_{25} k_{56} k_{61}}{k_{21} k_{16} k_{65} k_{52} } \right) = (-\Delta \mu_{\text{hyd}}) - f d_0
	\label{eq:AF}
\end{align}
and
\begin{align}
	\mathcal{A}_\mathcal{B} = k_B T \log \left( \frac{ k_{23} k_{34} k_{45} k_{51}}{k_{32} k_{25} k_{54} k_{43} } \right) = (-\Delta \mu_{\text{hyd}}) + f d_0. 
	\label{eq:AB}
\end{align}
The explicit forms of $J_{\mathcal{F}}$ and $J_{\mathcal{B}}$ as a function of $\{k_{ij}\}$ are available (see Eq. \ref{eq:JF}) but the expression is generally more complicated than the affinity. 
It is of note that at $f=0$, the chemical driving forces for $\mathcal{F}$ and $\mathcal{B}$ cycles are identical to be $-\Delta \mu_{\text{hyd}}$. 
The above decomposition of affinity associated with each cycle into the chemical driving force and the work done by the motor results from the Bell-like expression of transition rate between the states (2) and (5): $k_{25} = k_{25}^o e^{-\theta f d_0 / k_B T}, k_{52} = k_{52}^o e^{(1-\theta) f d_0 / k_B T}$ \cite{Liepelt07PRL,Fisher1999_PNAS,Fisher01PNAS} (see {\bf Materials and Methods}). 
From Eqs.~\ref{eq:dotQ}, ~\ref{eq:AF}, and ~\ref{eq:AB}, 
$\dot{Q}$ can be decomposed into the total free energy input ($\dot{E} = (J_\mathcal{F} + J_\mathcal{B} ) (-\Delta \mu_{\text{hyd}})$) and work production ($\dot{W} = (J_\mathcal{F} -  J_\mathcal{B} ) f d_0$). Hence,   
\begin{align}
	\dot{Q} &= (J_\mathcal{F} + J_\mathcal{B} ) (-\Delta \mu_{\text{hyd}}) - (J_\mathcal{F} -  J_\mathcal{B} ) f d_0 \nonumber\\
	&= \dot{E} - \dot{W}. 
\end{align}

A few points are noteworthy from the dependences of $V$, $D$, $\dot{Q}$, and $\dot{W}$ for kinesin-1 on $f$ and [ATP] (see Fig. \ref{fig_kinesin_6state}): 
(i) The stall condition, indicated with a white dashed line in each map, divided all the 2D maps of $V$, $D$, $\dot{Q}$, and $\dot{W}$ into two regions;  
(ii) In contrast to $V$ (Fig. \ref{fig_kinesin_6state}D) and $D$ (Fig. \ref{fig_kinesin_6state}E), which decrease monotonically with $f$, 
$\dot{Q}$ and $\dot{W}$ display non-monotonic dependence on $f$ (Figs. \ref{fig_kinesin_6state}F, G).
At high [ATP], $\dot{Q}$ is locally maximized at $f=10$ pN, whereas $\dot{W}$ is maximized at 5 pN and locally minimized at 10 pN (see Fig. \ref{fig_kinesin_6state}I calculated at [ATP]=2 mM).

At the stall force $f_\text{stall}$ ($\approx 7$ pN) (Figs. \ref{fig_kinesin_6state}B, \ref{fig_kinesin_6state}I, black dashed line), the reaction current of the $\mathcal{F}$-cycle is exactly balanced with that of $\mathcal{B}$-cycle ($J= J_\mathcal{F}- J_\mathcal{B} = 0$), giving rise to zero work production ($\dot{W} = f V = f d_0J= 0$). 
The numbers of forward and backward steps taken by kinesin motors are identical, and hence there is no net directional movement ($V=0$) \cite{Cross05Nature}.
Importantly, even at the stall condition, kinesin-1 consumes the chemical free energy of ATP hydrolysis, dissipating heat in both forward and backward steps, and hence rendering $\dot{Q}=(J_{\mathcal{F}}+J_{\mathcal{B }})(-\Delta \mu_{\text{hyd}})$ always  positive.

Next, branching of reaction current of the $\mathcal{F}$-cycle into $\mathcal{B}$-cycle with increasing $f$ gives rise to non-monotonic changes of $\dot{Q}$ and $\dot{W}$ with $f$.  
At small $f$, $\dot{Q}$ decreases with $f$ because exertion of load gradually deactivates the $\mathcal{F}$-cycle via the decrease of $\mathcal{A}_\mathcal{F}$ (Eq.\ref{eq:dotQ}, Eq.\ref{eq:AF}, Fig. \ref{fig_QWVD6_supple}C) and $J_\mathcal{F}$ (Fig. \ref{fig_QWVD6_supple}A). 
By contrast, $\mathcal{A}_\mathcal{B}$ and $J_\mathcal{B}$ increase with $f$ (Eq.\ref{eq:AB}, Figs. \ref{fig_QWVD6_supple}A and \ref{fig_QWVD6_supple}D), which leads to an increase of $\dot{Q}_{\mathcal{B}}$.
The non-monotonic dependence of $\dot{W}$ on $f$ can be analyzed in a similar way. 
The forward and backward currents along the cycles $\mathcal{F}$ and $\mathcal{B}$ (i.e., $J_\mathcal{F}$ and $J_\mathcal{B}$) are negatively correlated (Figs. \ref{fig_QWVD6_supple}A and \ref{fig_QWVD6_supple}B).

Motor head distortion at high external stress hinders the binding and hydrolysis of ATP in the catalytic site \cite{Uemura03NSB,Hyeon07PNAS,Hyeon11BJ}; $k_{ij}=0$ when ATP cannot be processed. 
This effect is modeled into the rate constants such that $k_{ij} = 2k_{ij}^o  (1 + e^{\chi_{ij} f d_0 /k_B T})^{-1}$ with $\chi_{ij} > 0$ ($k_{ij} \neq k_{25}, k_{52}$) \cite{Liepelt07PRL}. 
Thus, it naturally follows that $\dot{Q}=0$ when $f$ is much greater than $f_{\text{stall}}$ (Figs.\ref{fig_kinesin_6state}B, F, I). 

\begin{figure}[t]
	\centering
	\includegraphics[scale=0.42]{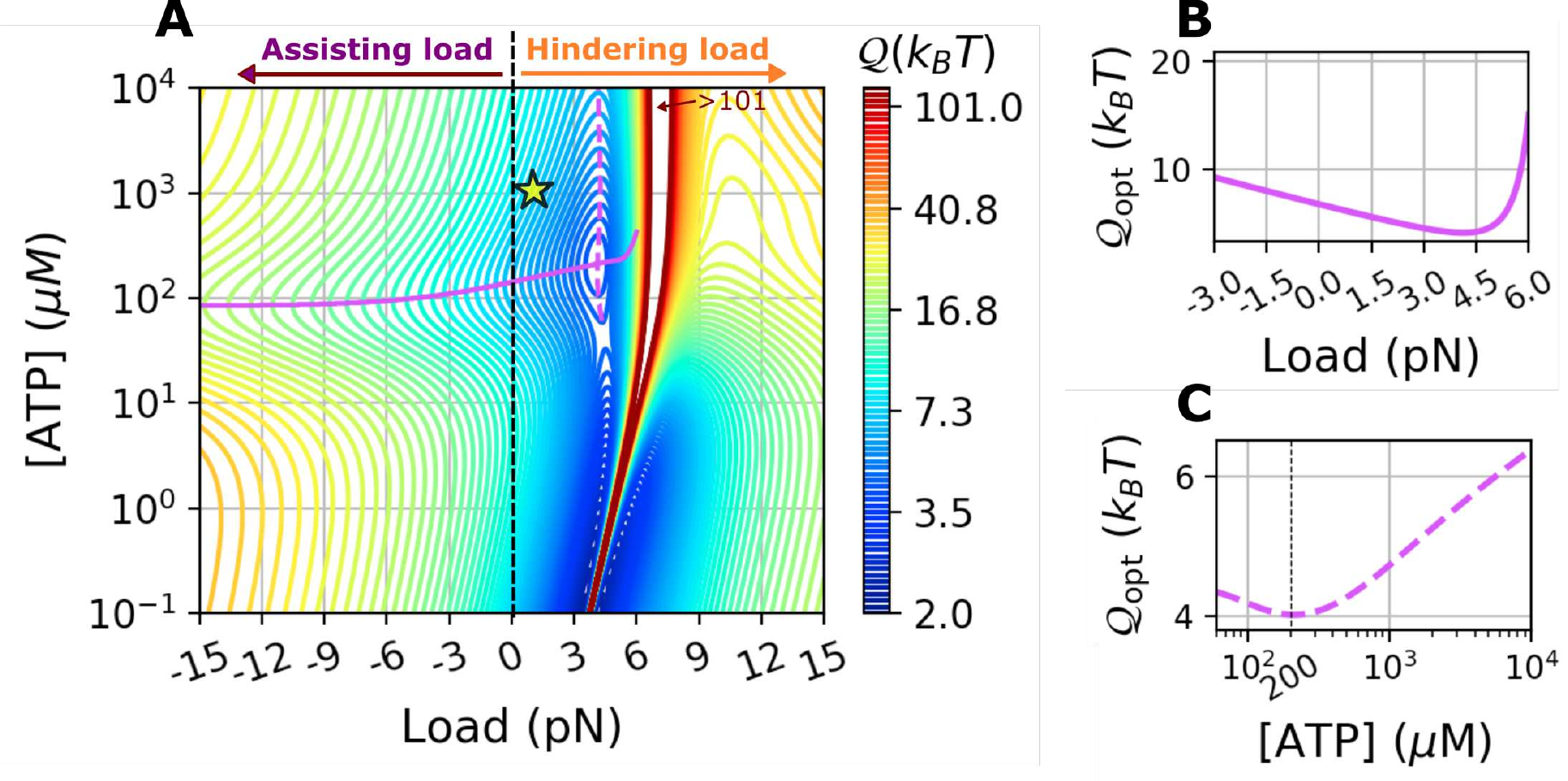}
	\caption{$\mathcal{Q}$ calculated based on kinesin-1 data \cite{Visscher99Nature} using the 6-state double-cycle model \cite{Liepelt07PRL} at varying $f$ and [ATP], where $f>0$ and $f<0$ signify the hindering and assisting load, respectively.  
		{\bf A.} 2-D contour plot of $\mathcal{Q} = \mathcal{Q}(f, [\text{ATP}])$. 
		A suboptimal point $\mathcal{Q}_{\text{opt}}\approx 4$ $k_BT$ is found at $f=4.1$ pN and $[\text{ATP}]\approx 210$ $\mu$M. 
		The solid lines in magenta are the loci of locally optimal $\mathcal{Q}$ at varying [ATP] for a given $f$ ({\bf B}).
		The dashed lines in magenta are the loci of locally optimal $\mathcal{Q}$ at varying $f$ for a given [ATP] ({\bf C}). 
		The star symbol indicates the cellular condition of $[\text{ATP}]\approx 1$ mM and $f\approx 1$ pN. 
	}		
	\label{fig_Qa_contour}
\end{figure}

\subsection{Quantification of $\mathcal{Q}$ for kinesin-1}
Unlike $V$, $D$, and $\dot{Q}$, which are maximized at large [ATP] and small $f$ (Figs. \ref{fig_kinesin_6state}D, E, F), 
the uncertainty measure $\mathcal{Q}(f, \text{[ATP]})$ displays a complex functional dependence (Fig. \ref{fig_Qa_contour}A). 
(i) Small $\mathcal{Q}$ at low [ATP] and $f$, which approaches the lower bound of 2$k_B T$, is a trivial outcome of the detailed balance condition where [ATP] is balanced with [ADP] and [P$_i$]. 
The motor, without chemical driving force and only subjected to thermal fluctuations ($\dot{Q}\rightarrow 0$), is on average motionless ($V\rightarrow 0$); $\mathcal{Q}$ is minimized in this case ($\mathcal{Q}\rightarrow  2k_BT$).
(ii) $\mathcal{Q}$ is generally smaller below the stall condition, $f<f_{\text{stall}}([\text{ATP}])$, demarcated by the white dashed lines in Fig.\ref{fig_kinesin_6state}. 
In this case, the reaction current along the $\mathcal{F}$-cycle is more dominant than that above the stall.  
At the stall, $\mathcal{Q}$ diverges because of $V\rightarrow 0$ and $\dot{Q}\neq 0$.  
(iii) Notably, a suboptimal value of $\mathcal{Q}\approx 4$ $k_BT$ is identified at [ATP] $= 210$ $\mu$M and $f = 4.1$ pN (Figs. \ref{fig_Qa_contour}). 
(iv) At $f \approx 4$ pN, $4 k_B T \lesssim \mathcal{Q} \lesssim 6 k_B T$ over the broad range of [ATP] ($=1$ $\mu$M$-$10 mM) remaining close to the local minimum value $4 k_B T$ (Fig. \ref{fig_Qa_contour}C), which indicates that kinesin-1 works robustly against the variation of [ATP] in the cell.

\begin{figure*}[ht]
	\centering
	\includegraphics[scale=0.6]{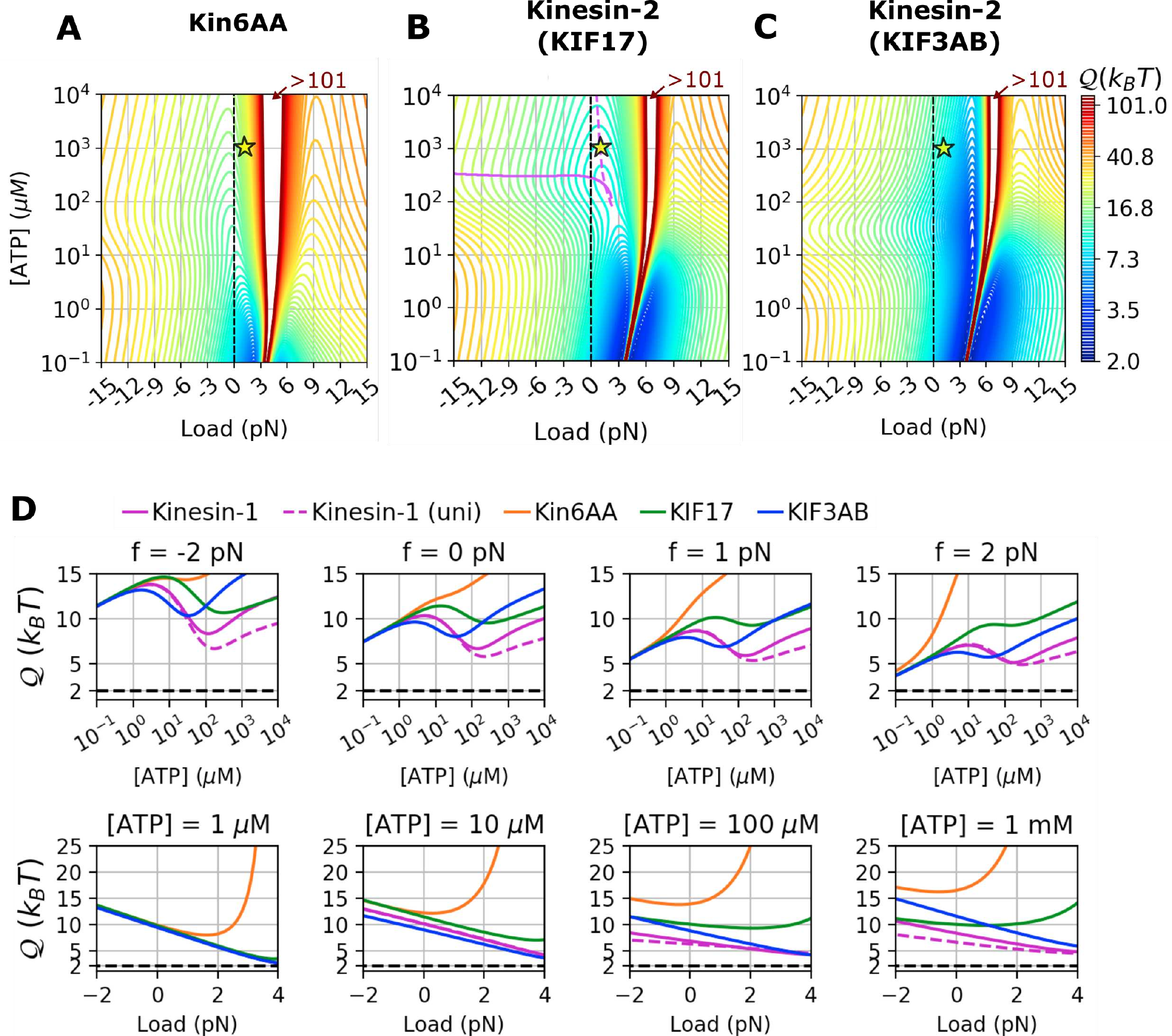}
	\caption{$\mathcal{Q}(f,[\text{ATP}]$ calculated for  
		({\bf A}) Kin6AA \cite{clancy2011NSMB}, 
		({\bf B}) homodirmeric kinesin-2 KIF17 \cite{Milic:2017:PNAS}, and 
		({\bf C}) heterotrimeric kinesin-2 KIF3AB \cite{Milic:2017:PNAS}.
		In {\bf A}-{\bf C}, the condition of $[\text{ATP}]\approx 1$ mM and $f\approx 1$ pN is indicated with the star symbols. 
		{\bf D}. $\mathcal{Q}(\text{[ATP]})$ at fixed $f$ (upper panels) and $\mathcal{Q}(f)$ at fixed [ATP] (lower panels) calculated for kinesin-1 (solid magenta lines for 6-state network model and dashed magenta lines for unicyclic model), Kin6AA (orange lines), KIF17 (green lines), and KIF3AB (blue lines).
		The black dashed lines depict $\mathcal{Q}=2k_BT$. 
	}
	\label{fig_Qa_other_kinesins}
\end{figure*}

\begin{figure*}[ht]
	\centering
	\includegraphics[scale=0.6]{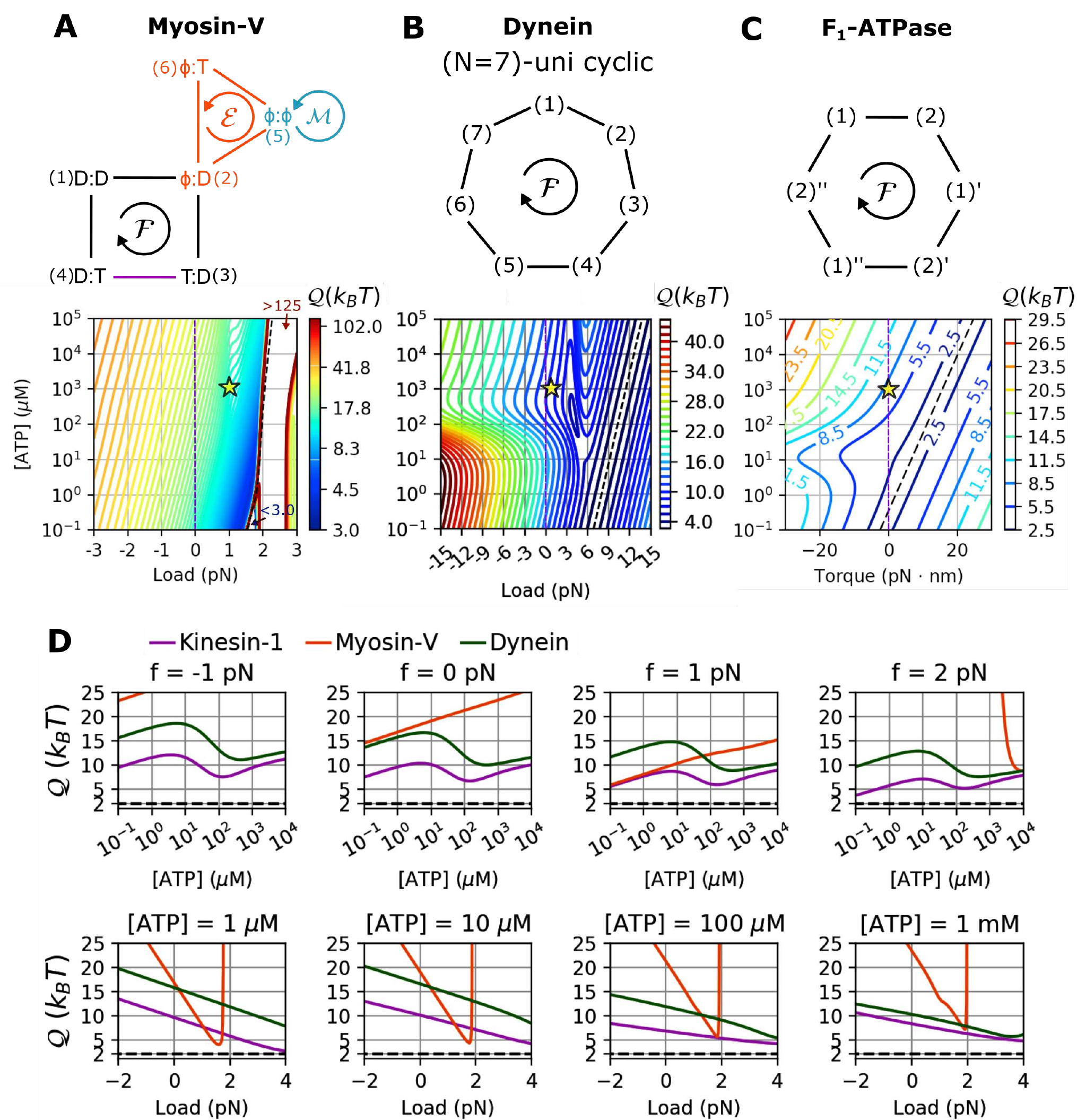}
	\caption{$\mathcal{Q}$ for various motors as a function of $f$ and [ATP]. 
		{\bf A.} Kinetic model for myosin-V consisting of three cycles $\mathcal{F}$, $\mathcal{E}$, and $\mathcal{M}$ \cite{Bierbaum:2011:BPJ}. 
		$\mathcal{Q}(f, [\text{ATP}])$ calculated at [ADP] = 70 $\mu$M and [P$_i$] = 1 mM (See also Fig. \ref{fig_myosin_various_70_1000}D for 2-D heat map).
		{\bf B.} $(N=7)$-unicyclic kinetic model for cytoplasmic dynein, and the corresponding $\mathcal{Q}(f, [\text{ATP}])$ calculated based on the kinetic parameters provided in Ref. \cite{Sarlah:2014:BPJ} at [ADP] = 70 $\mu$M and [P$_i$] = 1 mM (See also Fig.\ref{fig_dynein_various}D for 2-D heat map).
		{\bf C}. $\mathcal{Q}(f, [\text{ATP}])$ at [ADP] = 70 $ \mu$M and [P$_i$] = 1 mM (See also Fig.\ref{fig_dynein_various}D for 2-D heat map) using the kinetic model for F$_1$-ATPase from Ref. \cite{Gerritsma:2010}. 
		Other quantities such as $V$, $D$, and $\mathcal{A}$ as a function of $f$ and [ATP] are provided in Figs. \ref{fig_myosin_various_70_1000}, \ref{fig_dynein_various}, \ref{fig_f1_various}.
		{\bf D}. $\mathcal{Q}(\text{[ATP]})$ at fixed $f$ (upper panels) and $\mathcal{Q}(f)$ at fixed [ATP] (lower panels) for kinesin-1 (magenta), myosin-V (orange), and dynein (green).
	}
	\label{fig_Qa_other_motor_proteins}
\end{figure*}

\subsection{Comparison of $\mathcal{Q}$ between different types of kinesins}
The dynamic property of molecular motor differs from one motor type to another. 
Effect of modifying motor structure on the transport properties as well as on the directionality and processivity of molecular motor has been of great interest because it provides glimpses into the design principle of a motor at molecular level \cite{liao2009JMB,Bryant2007PNAS,Hyeon11BJ,Jana2012PLoS,hinczewski2013PNAS2}. 
To address how modifications to motor structure alter the transport efficiency of motor, we analyze single-molecule motility data of a mutant of kinesin-1 (Kin6AA), and homodimeric and heterotrimeric kinesin-2 (KIF17 and KIF3AB). 

Data of Kin6AA, a mutant of kinesin-1 that has a longer neck-linker domain, were taken from Ref. \cite{clancy2011NSMB}.
Six amino-acid residues inserted to the neck-linker reduce the internal tension along the neck-linker which plays a critical role for regulating the chemistry of two motor heads and coordinating the hand-over-hand motion \cite{Hyeon07PNAS}. 
Disturbance to this motif is expected to affect $V$ and $D$ of the wild type.
We analyzed the data of Kin6AA again using the 6-state network model (Figs. \ref{fig_kinesin_6state}A, \ref{fig_exp_Kin6AA}, \ref{fig_Kin6AA_various}, Table \ref{table}. See {\bf SI} for detail), indeed finding reduction of $V$ and $D$ (Fig. \ref{fig_Kin6AA_various}A) as well as its stall force (Fig. \ref{fig_Kin6AA_various}A, white dashed line).
Of particular note is that the rate constant $k_{25}$ associated with the mechanical stepping process is reduced by two order of magnitude (Table \ref{table}).
In $\mathcal{Q}(f,[\text{ATP}])$ (Fig. \ref{fig_Qa_other_kinesins}A), 
the suboptimal point observed in kinesin-1 (Fig. \ref{fig_Qa_contour}A) vanishes (Fig. \ref{fig_Qa_other_kinesins}A), and $\mathcal{Q}$ diverges around $\sim$ 4 pN due to the decreased stall force.
Finally, overall, the value of $\mathcal{Q}$ has increased dramatically. 
This means that compared with that of kinesin-1 ($\mathcal{Q}\approx 7$ $k_BT$), the trajectory of Kin6AA is less regular and unpredictable ($\mathcal{Q}\approx 20$ $k_BT$) at $f=1$ pN and [ATP] = 1 mM, that roughly represents the cellular condition \cite{Welte:1998,Shubeita:2008}. 
Thus, Kin6AA is three fold less efficient than the wild-type in cargo transport.  

Next, the values of $\mathcal{Q}$ were calculated for two active forms of vertebrate kinesin-2 class motors responsible for intraflagellar transport (IFT). 
KIF17 is a homodimeric form of kinesin-2, and KIF3AB is a heterotrimeric form made of KIF3A, KIF3B, and a nonmotor accessory protein, KAP. 
To quantify their motility properties, we digitized single-molecule motility data  from Ref. \cite{Milic:2017:PNAS} and fitted them to the 6-state double-cycle model (Figs. \ref{fig_exp_KIF17}, \ref{fig_exp_KIF3AB}) (See {\bf SI} for detail).
$\mathcal{Q}(f,[\text{ATP}])$'s of KIF17 and KIF3AB are qualitatively similar to that of kinesin-1 with some variations. 
$\mathcal{Q}$ for KIF17 forms a shallow local minimum of $\mathcal{Q}\approx 9.2$ $k_BT$ at [ATP] = 200 $\mu$M and $f = 1.5$ pN (Fig. \ref{fig_Qa_other_kinesins}B), whereas 
such suboptimal condition vanishes in KIF3AB (Fig. \ref{fig_Qa_other_kinesins}C). 
KIF3AB, however, display a local valley of $\mathcal{Q}$ around $f \sim $ 4 pN and $1$ $\mu$M $\lesssim \text{[ATP]}  \lesssim 10$ mM in which $\mathcal{Q} \approx 4 k_B T$.

The plots of $\mathcal{Q} (\text{[ATP]})$ at fixed $f$ and $\mathcal{Q} (f)$ with fixed [ATP] in Fig. \ref{fig_Qa_other_kinesins}D recapitulate the difference between different classes of kinesins more clearly. 
The following features are noteworthy.
(i) An extension of neck-linker domain (Kin6AA, orange lines) dramatically increases $\mathcal{Q}$ compared with the wild type (Kinesin WT, magenta lines). 
(ii) Non-monotonic behaviors of $\mathcal{Q} (\text{[ATP]})$ are qualitatively similar for all kinesins 
although $\mathcal{Q}$ is, in general, the smallest for kinesin-1.
(iii) The movement of KIF3AB (black lines) becomes the most regular at low [ATP] ($ \lesssim 10$ $\mu$M). 
(iv) $\mathcal{Q}$ for kinesin-1 analyzed using the ($N$=4)-state unicyclic model (dashed magenta lines) displays only small deviations as long as $0\lesssim f\lesssim 4$ pN $\ll f_{\text{stall}} \approx 7$ pN.

\subsection{Comparison of $\mathcal{Q}$ among different types of motors} 
We further investigate $\mathcal{Q}(f, \text{[ATP]})$ for other motor types, myosin-V, dynein, and F$_1$-ATPase using the kinetic network models proposed in the literature \cite{Bierbaum:2011:BPJ,Sarlah:2014:BPJ,Gerritsma:2010}.

{\it Myosin-V}: 
The model studied in Ref. \cite{Bierbaum:2011:BPJ} consists of chemomechanical forward cycle $\mathcal{F}$, dissipative cycle $\mathcal{E}$, and pure mechanical cycle $\mathcal{M}$ (Fig. \ref{fig_Qa_other_motor_proteins}A).
In $\mathcal{F}$-cycle, myosin-V either moves forward by hydrolyzing ATP or takes backstep via ATP synthesis.
In $\mathcal{M}$-cycle, myosin-V moves backward under the load without involving chemical reactions. 
The $\mathcal{E}$-cycle, consisting of ATP binding [$(2) \rightarrow (5)$], ATP hydrolysis [$(5) \rightarrow (6)$], and ADP release [$(6) \rightarrow (2)$] (Fig. \ref{fig_Qa_other_motor_proteins}B), was originally introduced to connect the two cycles $\mathcal{F}$ and $\mathcal{M}$.
The calculation of $J_\mathcal{E}$ and $J_\mathcal{F}$ reveals that a gradual deactivation of $\mathcal{F}$-cycle with decreasing [ATP] activates the $\mathcal{E}$-cycle (Fig. \ref{fig_myosin_various_70_1000}A, $100$ $ \mu$M $\lesssim \text{[ATP]} \lesssim 1$ mM,  $f \lesssim 1$ pN). 
Thus, $\mathcal{E}$-cycle can be regarded a futile $\mathcal{F}$-cycle, which is activated when chemical driving force is balanced with a load $f$ at low [ATP]. 
$\mathcal{Q}( f,\text{[ATP]})$ calculated at [ADP] = 70 $\mu$M and [P$_i$] = 1 mM using the rate constants from Ref. \cite{Bierbaum:2011:BPJ} (see \textbf{SI} for details and Fig. \ref{fig_myosin_various_70_1000}) reveals no local minimum in this condition. 
However, at [ADP] = 0.1 $\mu$M and [P$_i$] = 0.1 $\mu$M, which is the condition used in Ref. \cite{Bierbaum:2011:BPJ},  a local minimum with $\mathcal{Q}$ = 6.5 $k_B T$ is identified at $f =$ 1.1 pN and  [ATP] = 20 $\mu$M  (Fig. \ref{fig_myosin_various}D, Table \ref{table_opt}).
Both values of $f$ and [ATP] at the suboptimal condition of myosin-V are smaller than those of kinesin-1 (Table \ref{table_opt}).
In (N=2)-unicyclic model for myosin-V (Fig. \ref{fig_myosin_uni}) \cite{Kolomeisky:2003:BPJ}, $\mathcal{Q}$ has local valley around $f\sim 2$ pN and [ATP] $\sim 10$ $\mu$M. 
Similar to the result from the multi-cyclic model with [ADP] = 0.1 $\mu$M and [P$_i$] = 0.1 $\mu$M, 
the values of $f$ and [ATP] along the valley of $\mathcal{Q}$ are smaller than the values optimizing $\mathcal{Q}$ for the kinesin-1 (Table \ref{table_opt}).\\

{\it Dynein}:
Dynein is a family of $(-)$-end directed cytoskeletal motor.
There are two groups of dyneins: cytoplasmic and axonemal dyneins. 
Cytoplasmic dyneins involve the transport of cellular cargoes whereas axonemal dyneins are responsible for generating the beating motion of cilia or flagella by sliding microtubles in the axonemes. Here we study cytoplasmic dyneins whose locomotion along microtubules is pertinent to the issue discussed here.  

$\mathcal{Q}(f,[\text{ATP}])$ for cytoplasmic dyneins was evaluated by considering (N=7)-unicyclic kinetic model (Fig. \ref{fig_Qa_other_motor_proteins}B) based on a previous study \cite{Sarlah:2014:BPJ}.
The original model (Fig. 5A of Ref. \cite{Sarlah:2014:BPJ}) describes the major pathway of tightly coupled dimeric dynein whose linker connecting the head domains of two dynein monomers is short and stiff. 
This major pathway is found in Ref. \cite{Sarlah:2014:BPJ} from the kinetic simulation of their elastomechanical model whose the transitions between chemical states and the mechanical movements of motors are described by elastic-energy- and load-dependent rate constants.
Although the futile cycle, which branches out of the major pathway, is expected at large hindering loads \cite{Sarlah:2014:BPJ}, we consider a simpler unicycle model; as shown in kinesin-1 (Fig. \ref{fig_Qa_other_kinesins}D), as long as $f$ is small ($f<f_c$), the system under the unicycle model behaves similarly to a more complicated model. 
The model consists of 7 states: dissociation of Pi [$(1) \ra (2)$]; dissociation of ADP  [$(2)\ra(3)$]; ATP binding [$(3) \ra (4)$]; dissociation of microtuble binding domain (MTBD) from the filament [$(4) \ra (5)$]; power stroke [$(5) \ra (6)$]; linker swinging to the pre-power stroke state [$(6) \ra (7)$]; MTBD binding to the filament [$(7) \ra (1)$].
We also assume only the rate constants describing the mechanical transition of dynein depend on $f$. 
The more detailed description of the model is given in {\bf SI}.
$\mathcal{Q}(f,[\text{ATP}])$ calculated from the model is locally minimized to $\mathcal{Q}\approx 5.2 ~ k_B T$ at $f$ = 3.9 pN, [ATP] = 200 $\mu$M (Fig. \ref{fig_Qa_other_motor_proteins}D, \ref{fig_dynein_various}D). 
This condition of local minimum is compatible with that of kinesin-1 (Table \ref{table_opt}).
\\

{\it F$_1$-ATPase}:
F$_1$-ATPase is a rotary molecular motor. In vivo, it combines with F$_0$ subunit and synthesizes ATP by using proton gradient across membrane.
$\mathcal{Q}(\tau,[\text{ATP}])$ calculated using the (N=2)-state unicyclic model in Ref. \cite{Gerritsma:2010} (also see {\bf SI} for the detailed description of the model) reveals that 
there is a valley around torque $\tau \approx -10$ pN$\cdot$nm and [ATP] $\approx 10$ $\mu$M reaching $\mathcal{Q} \approx 4$ $k_B T$ (Fig. \ref{fig_Qa_other_motor_proteins}C).
Notably, $\mathcal{Q}$ for F$_1$-ATPase is optimized at hindering load ($\tau <0$) in which ATP is synthesized, which comports well with the biologically known role of F$_1$-ATPase as an ATP synthase \emph{in vivo}.
\\

To highlight the difference between the motors, 
we plot $\mathcal{Q} (\text{[ATP]})$ at fixed $f$ and $\mathcal{Q} (f)$ at fixed [ATP] in Fig. \ref{fig_Qa_other_motor_proteins}D, which find $\mathcal{Q}_{\text{kinesin-1}} < \mathcal{Q}_{\text{dynein}} < \mathcal{Q}_\text{myosin-V}$ over the broad range of $f$ and [ATP]. 
We note that at a special condition ([ATP] ($=1-10$ $\mu$M) and $f$ ($=1 - 2$ pN)) $\mathcal{Q}_\text{myosin-V}$ is smaller than the values of other motors.

\begin{figure}[t]
	\centering
	\includegraphics[scale=1]{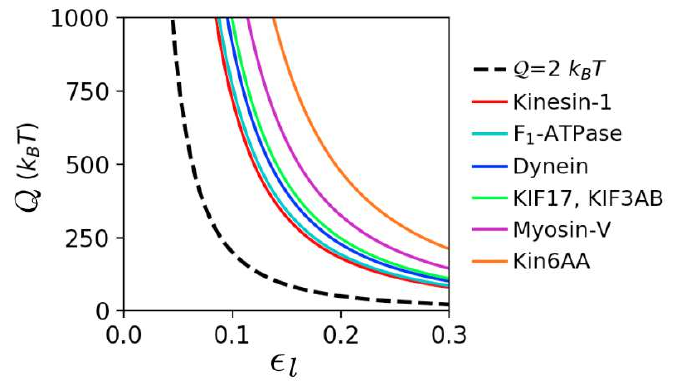}
	\caption{Plots of $Q$ versus $\epsilon_l(=\sqrt{\langle (\delta l)^2\rangle}/\langle l\rangle)$ for various motors under the cellular condition, [ATP] = 1 mM \cite{Milo:2015} and $f = 1$ pN \cite{Welte:1998,Shubeita:2008}.  
		The plot for the ideal motor ($\mathcal{Q}=2$ $k_BT$) is plotted with a dashed line. 
		$\mathcal{Q}/k_BT=7.2$ (kinesin-1), 7.7 (F$_1$-ATPase), 9.1 (dynein), 
		9.9 (KIF17), 9.9 (KIF3AB), 13 (myosin-V), 19 (Kin6AA). 
		For F$_1$-ATPase, zero torque and [ATP] $=1$ mM are assumed as the cellular condition. 
	}
	\label{fig_Qa_scheme}
\end{figure}

\section{Discussion} 
Biological motors are far superior to macroscopic machines in harnessing free energy into linear movement. 
The thermal noise is utilized to rectify the ATP binding/hydrolysis-coupled conformational dynamics into unidirectional movement,    
which conceptualizes the Brownian ratchet \cite{Astumian97Science}, but it also comes with a cost of overcoming the thermal noise that makes the movement of biological motors inherently stochastic and error-prone. 
Mechanism of harnessing energy into faster and more precise motion is critical for the accuracy of cellular computation. 
The uncertainty measure $\mathcal{Q}$ assesses the efficiency of improving the speed and regularity of dynamics for a given energetic cost.

Here, we have quantified the uncertainty measure $\mathcal{Q}$ for various biological motors. 
We found that the values of $\mathcal{Q}$ for various motors are all semi-optimized near the cellular condition (star symbols marking $f\approx 1$ pN and [ATP]$=1$ mM in Figs.~\ref{fig_Qa_contour}, ~\ref{fig_Qa_other_kinesins}, ~\ref{fig_Qa_other_motor_proteins}).
$Q$ versus $\epsilon_l$ plots (Fig.\ref{fig_Qa_scheme}, $\epsilon_{\theta}$ for F$_1$-ATPase motor) for the various motors, sorting the motors in the increasing order of $\mathcal{Q}$, and their lower bound dictated by $\mathcal{Q}=Q\epsilon_l^2=2k_BT$
are reminiscent of the recent study on the free-energy cost of accurate biochemical oscillations \cite{Cao2015NatPhys}. 
The plots indicate that kinesin-1 is the best motor whose $\mathcal{Q}$($\approx 7.2$ $k_BT$) approaches the bound of ideal case ($\mathcal{Q}=2$ $k_BT$).
Note that $\mathcal{Q}(\approx 19$ $k_BT$) for the mutant kinesin-1 (Kin6AA) is significantly greater than that for the wild-type.   
The structure of $\mathcal{Q}(f,[\text{ATP}])$ and the suboptimal condition of $\mathcal{Q}$ differ from one motor type to another. 

\begin{figure}[t]
	\centering
	\includegraphics[scale=0.9]{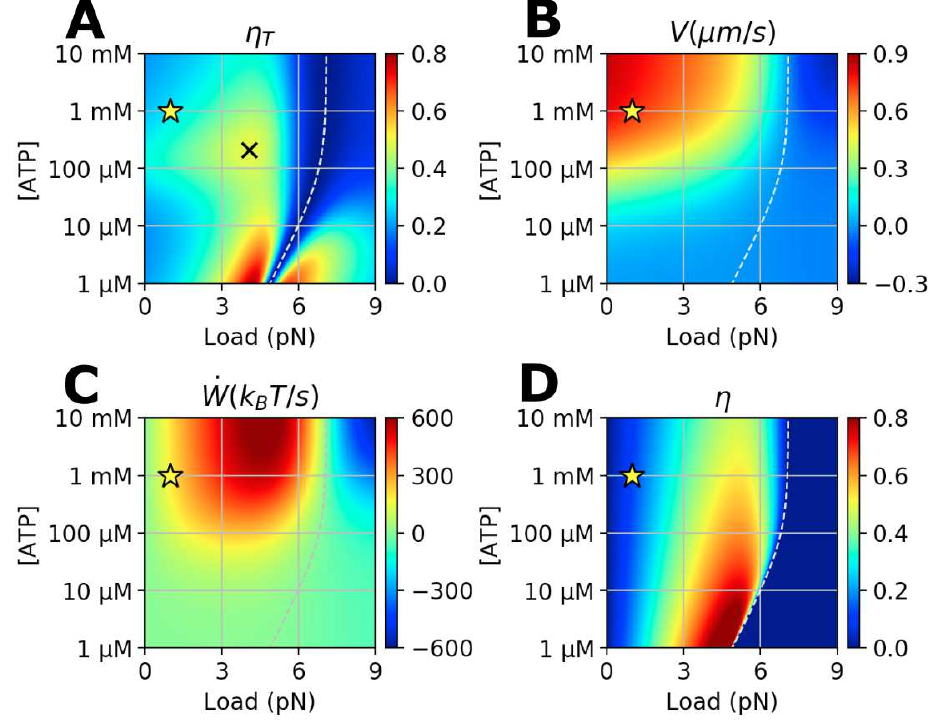}
	\caption{Various quantities calculated for kinesin-1 at varying conditions of $f$ and [ATP]. 
		{\bf A}. Transport efficiency $\eta_T(=2k_BT/\mathcal{Q})$. 
		A suboptimal point $\eta_T^*\approx 0.48$ (indicated by $\times$) is formed at $f=4.1$ pN and $[\text{ATP}]=210$ $\mu$M. {\bf B}. Transport speed $V(f,[\text{ATP}])$, {\bf C}. Work production $\dot{W}(f,[\text{ATP}])$, 
		{\bf D}. Power efficiency calculated using $\eta \equiv \dot{W}/\dot{E}$. 
		For $f > f_\text{stall}$, we set $\eta=0$ for convenience because the motor moves backward and $\dot{W}<0$. 
		At the cellular condition ($f\approx 1$ pN and [ATP] $\approx 1$ mM), indicated by the star symbol in each panel, $\eta_T = 0.28$, $V = 0.74$ $\mu$m/s, $\dot{W} = 182$ $k_B T/s$, and $\eta = 0.12$. 
	}  
	\label{efficiency}
\end{figure} 

Minimizing $\mathcal{Q}$ towards its lower bound $2k_BT$ for optimal transport is equivalent to maximizing 
the transport efficiency, which can be defined as \cite{dechant2017Arxiv}   
\begin{align}
	\eta_T(f,[\text{ATP}])=\frac{2k_BT}{\mathcal{Q}(f,[\text{ATP}])},    
\end{align}
where $\eta_T$ is bounded in the interval $0\leq\eta_T\leq 1$. 
It is of particular note that the structure of $\eta_T(f,[\text{ATP}])$ (equivalently $\mathcal{Q}(f,[\text{ATP}])$) differs significantly from that of other quantities such as the flux $J(f,[\text{ATP}])$ \cite{Brown:2017:PNAS} (equivalent to $V(f,[\text{ATP}])$),  work production (power) $\dot{W}(f,[\text{ATP}])$, 
and the power efficiency $\eta(f,[\text{ATP}])\equiv \dot{W}(f,[\text{ATP}])/\dot{E}(f,[\text{ATP}])$ (Fig.\ref{efficiency} for kinesin-1. See \textbf{SI} Figures for other motors). 
Remarkably, only $\eta_T(f,[\text{ATP}])$, not $V$, $\dot{W}$, nor $\eta$, displays a suboptimal peak near the cellular condition. 

To what extent can our findings on the \emph{in vitro} single motor properties be generalized into those in live cells?
First,  the force hindering the motor movement varies with cargo size and subcellular location; the load or viscoelastic drag exerted against motors inside the cell varies dynamically \cite{Narayanareddy:2014,Wortman:2014}.
Yet, actual forces opposing the cargo movement in cytosolic environment are $\lesssim$ 1 pN \cite{Welte:1998,Shubeita:2008}. 
Since $\mathcal{Q}$'s for microtubule-binding motors, kinesin-1, kinesin-2, and dynein, are narrowly tuned, varying only a few $k_B T$ over the range of $0 \leq f \leq 4$ pN at [ATP] = 1 mM (Figs. \ref{fig_Qa_other_kinesins}D, \ref{fig_Qa_other_motor_proteins}D), 
our discussion on the \emph{in vitro} single motor property can be extended to the cargo transport in cytosolic environment. 
Next, a team of motors is often responsible for cargo transport in the cell \cite{li2016BJ}. 
It has, however, been shown that the extent of coordination between two kinesin motors attached to a cargo is not significant under low load and saturating ATP \cite{carter2008BJ}.
Although trajectories generated by multiple motors have not been analyzed here, 
extension of the present analysis to such cases is straightforward.

In the axonal transport, of particular importance is the fast and timely delivery of cellular material, the failure of which is linked to neuropathology \cite{mandelkow2002TCB,Hafezparast2003Science}. 
Since there are already numerous regulatory control mechanisms as well as other motors, it could be argued that the role played by the optimized single motor transport efficiency is redundant in light of the overall function of axonal transport. Yet, given that cellular regulations are realized through multiple layers of checkpoints \cite{AlbertsBook}, the optimized transport efficiency of motors at single molecule level can also be viewed as one of the checkpoints that assure the optimal cargo transport.  

Taken together, the thermodynamic uncertainty relation, a general principle for dissipative processes in nonequilibrium steady states, offers quantitative insight into the energy-speed-precision trade-off relation for biological systems. 
Here, we have adapted this principle to assess the transport efficiency of biological motors in terms of $\mathcal{Q}$. 
With a multitude of time traces of a biological motor generated at varying conditions at hand, it is straightforward to calculate the uncertainty measure $\mathcal{Q}$ as well as other dynamic quantities of interest by mapping the dynamics of the motor to an adequate kinetic network. 
Given that there are many possible directions involving the design principle of biological motors, it is significant to find that biological motors  indeed possess a semi-optimal transport efficiency under the cellular condition. 
Finally, it is of great interest to extend the proposed concept and analysis using $\mathcal{Q}$ to other energy-consuming biological processes.

\section{Materials and Methods}
To calculate the uncertainty measure $\mathcal{Q}$ of a motor, we first define a chemical network model 
that can describe the dynamics of the motor in terms of a set of rate constants $\{k_{ij}\}$, where $k_{ij}(f, [\text{ATP}])$ is the transition rate from the $i$-th to $j$-th state and depends on both load $f$ and [ATP].
Next, we fit the experimental data of $V$ and $D$ obtained under varying conditions of $f$ and [ATP] to the formal expressions of $V(\{k_{ij}(f,[\text{ATP}])\})$ and $D(\{k_{ij}(f,[\text{ATP}])\})$ \cite{Koza1999,Lebowitz:1999,barato2015PRL} (the details of the procedure is described in the next paragraph and in {\bf SI}). 
The fit determines the set of rate constants $k_{ij}(f, [\text{ATP}])$ \cite{Hwang2017JPCL}, and allows us to calculate the reaction current ($J$), current fluctuation ($\delta J^2$), affinity 
($\mathcal{A}$, net driving force),  heat dissipation ($\dot{Q}$), and hence $\mathcal{Q}$ (Eq.\ref{eqn:inequality2}) associated with the network.

To define the motor's mechanical step using the transitions in chemical state space, we assign a set of distance metric $\{d_{ij}\}$ to each transition.
For example, if the transition from the $i$-th to $j$-th state is made at time $t$, 
the increment of displacement is $l(t+dt) = l(t) + d_{ij}$, where we assign nonzero value $d_{ij}\neq 0$ ($d_{ij} = - d_{ji}$) for the transition associated with physical movement along the filament, and $d_{ij}=0$ for pure chemical transitions. 
Next, to calculate the probability density of motor along the spatial and chemical state space we extend the method of generating function used in ref.~\cite{Koza1999} (see {\bf SI} for details).
The time evolution of the generating function described in terms of the generalized variable $z$ is described by the transition matrix 
\begin{equation} \label{eq:Gamma_manu} 
	\Gamma_{i j}(z) = 
	\begin{cases}
		k_{i j} e^{z d_{i j}} \text{, for $i \neq j$}\\
		- \sum_{m=1(\neq i)}^N k_{i m}\text{, for $i = j$}. 
	\end{cases}
\end{equation}
$\Gamma_{i j}(0)$ is a usual transition matrix for a master equation.
$V$ and $D$, defined at the asymptotic limit ($t \gg 1$), can be obtained from the derivatives of the largest eigenvalue $\lambda_0(z)$ of $\Gamma(z)$ \cite{Koza1999} (see {\bf SI}): 
\begin{align}
	V &= \lambda_0'(0)
\end{align}
and 
\begin{align}
	D &= \frac{\lambda_0''(0)}{2},
\end{align}
where the prime denotes a partial derivative with respect to $z$. 
This method can be employed to calculate the transport property associated with a subcycle of an arbitrarily complex chemical network (see {\bf SI}).


For kinesin-1, we considered a 6-state double-cycle kinetic network (Fig.\ref{fig_kinesin_6state}A). 
The load dependence of kinetic rate was modeled using   
$k_{25}(f) = k_{25}^o e^{- \theta f d_0 / k_B T}$ and $k_{52}(f) = k_{52}^o e^{(1-\theta) f d_0 / k_B T}$ 
for the mechanical step between the states (2) and (5), and $k_{ij} (f) = 2 k_{ij}^o (1 + e^{\chi_{ij} f d_0 / k_B T})^{-1}$ for other steps associated with ATP chemistry ($ij\neq 25, 52$).
The condition of $\chi_{ij} = \chi_{ji}$ makes only the mechanical transition contribute to the work \cite{Liepelt07PRL}.
The rate constants determined for the $\mathcal{F}$-cycle are copied to the corresponding chemical steps in the $\mathcal{B}$-cycle \cite{Liepelt07PRL}. 
For example, ADP dissociation rate constant $k_{23}$ of $\mathcal{B}$-cycle is equal to $k_{56}$ which describes ADP dissociation in $\mathcal{F}$-cycle. Similarly, 
$k_{32} = k_{65}, k_{34} = k_{61}, k_{43} = k_{16}, k_{45}(=k_{45}^{bi}[\text{ATP}]) =k_{12}(= k_{12}^{bi}[\text{ATP}]), \chi_{23} = \chi_{56}, \chi_{34} = \chi_{61}, \chi_{45} = \chi_{12}$.
Since the ATP hydrolysis free energy that drives the $\mathcal{F}$- and $\mathcal{B}$-cycle is identical, $ (k_{12} k_{25} k_{56} k_{61} / k_{21} k_{52} k_{65} k_{16} )= (k_{23} k_{34} k_{45} k_{52} / k_{32} k_{43} k_{54} k_{25})$; thus 
$k_{54} (= k_{21} ( k_{52}/k_{25} )^2)$ \cite{Liepelt07PRL}. 
Because of the paucity of data at high load condition that activates the $\mathcal{B}$-cycle, it is not easy to determine all the parameters for $\mathcal{F}$ and $\mathcal{B}$ cycles simultaneously using the existing data \cite{Visscher99Nature}. 
To circumvent this difficulty, we fit the data using the following procedure. 
First, the affinity $\mathcal{A}$ at $f=0$ was determined from our previous study that employed the (N=4)-state unicyclic model \cite{Hwang2017JPCL}. 
Even though $\mathcal{B}$-cycle is not considered in Ref. \cite{Hwang2017JPCL}, $J_\mathcal{B} \approx 0$ at $f \sim 0$, which justifies the use of unicyclic model at $f \ll f_{\text{stall}}$. 
Next, the range of parameters were constrained during the fitting procedure (Table \ref{table_fit}) based on the values obtained in \cite{Hyeon09PCCP,Liepelt07PRL}).
To fit the data globally, we employed the \texttt{minimize} function with `L--BFGS--B' method from the scipy library.

For Kin6AA, KIF17, and KIF3AB, motility data digitized from Ref. \cite{clancy2011NSMB,Milic:2017:PNAS} were fit to the same 6-state double-cycle network model used for kinesin-1.
For myosin-V, dynein, and F$_1$-ATPase, we employed kinetic network models and corresponding rate constants used in Ref. \cite{Bierbaum:2011:BPJ,Sarlah:2014:BPJ,Gerritsma:2010}. Further details are provided in {\bf SI}.
\\

{\bf Acknowledgements.} We thank Steven P. Gross for insightful comments on cargo transport in live cells. 
This work was performed in part at Aspen Center for Physics, which is supported by National Science Foundation grant PHY-1607611. 
We acknowledge the Center for Advanced Computation in KIAS for providing computing resources. 

\clearpage

\setcounter{figure}{0}  
\setcounter{equation}{0}
{\Large {\bf Supplementary Information} }

\renewcommand{\thefigure}{S\arabic{figure}} 
\renewcommand{\theequation}{S\arabic{equation}}

\section{Calculation of $V$ and $D$ of kinesin-1 in 6-state multi-cyclic model}
To obtain the expression of $V$ and $D$ for multicyclic kinetic network model  in terms of a set of rate constants $\{k_{ij}\}$, 
we have generalized the technique by Koza \cite{Koza1999} (Alternatively, technique based on the large deviation theory can be used. See Ref. \cite{Wachtel:2015:PRE,Lebowitz:1999}).
We define the generating functions for the given network model. 

In the 6-state double-cycle kinetic network (Fig. \ref{fig_kinesin_6state}A), we define the three distinct generating functions for $\F$, $\mathcal{B}$, and $\mathcal{X}$ cycles. 
The two generating functions for the subcycles, $\mathcal{F}$ and $\mathcal{B}$-cycles, are convenient to calculate the chemical current $J_\F$ and $J_{\mathcal{B}}$ in each subcycle. 
To calculate $V$ and $D$ in a convenient way, we have defined another generating function for $\mathcal{X}$-cycle, which is not explicit in the kinetic scheme in Fig. \ref{fig_kinesin_6state}A.  
The $\mathcal{X}$-cycle differs from $\F$, $\mathcal{B}$-cycle in that the former explicitly considers the physical location of the motor along the 1D track. 
Although $V$ is obtained either evaluating $V=d_0(J_\F-J_{\mathcal{B}})$ or $V=d_0J_{\mathcal{X}}$, it is not straightforward to decompose the diffusivity of motor $D$ into the contributions from $\F$ and $\mathcal{B}$-cycle.  

The expressions of $J_\F(\{k_{ij}\})$, $J_{\mathcal{B}}(\{k_{ij}\})$, $V (\{k_{ij}\})$ and $D(\{k_{ij}\})$ can be obtained by considering an asymptotic limit ($t \ra \infty$) of the corresponding generating function.  

In what follows, we provide the derivation of generating function in details. In order to derive the generating function, we introduce a generalized index for reaction cycle $\I$, with $\I=\F$, $\mathcal{B}$, or $\mathcal{X}$. 
\\

\subsection{ Master equation }
For a system with $N$ chemical states ($\{1, 2, \cdots, N\}$), a generalized state $\mu^\I(t)$ is defined by using the chemical state of the motor at time $t$ and the number of completed $\I$-cycles ($n_c^\I(t)$).
For kinesins whose dynamics can be mapped onto the 6-state double-cycle kinetic network model, 
if the motor is in the $i$-th chemical state ($i\in \{1, 2, \cdots, N\}$ with $N=6$) at time $t$, 
the generalized state of the motor in the $\I$-cycle is $\mu^\I(t) = i + N \times n_c^\I (t)$, where 
$\I$ could denote either $\F$, $\mathcal{B}$, or $\mathcal{X}$ depending on reader's interest. 
$P(\mu^\I, t)$ that represents the probability of the system being in $\mu^\I$ at time $t$, satisfies
\begin{widetext}
	\begin{equation}
		\begin{aligned}
			\frac{ \partial P(\mu^\I, t) } {\partial t} = \sum_{\xi}  K_{\mu^\I-\xi, \mu^\I} P( \mu^\I -\xi, t ) - K_{\mu^\I, \mu^\I-\xi} P(\mu^\I, t),
			\label{eq:master}
		\end{aligned}
	\end{equation}
\end{widetext}
where $K_{\mu, \nu} = \sum_{\alpha} k^\alpha_{\mu, \nu}$ and 
$k^\alpha_{ij}$ denotes the rate of transition from state $i$ to state $j$ that follows the $\alpha$-th pathway.
Here, the periodicity of network model imposes $k^\alpha_{\mu+N, \nu+N} = k^\alpha_{\mu, \nu}$, $K_{\nu + N, \mu+N} = K_{\nu, \mu}$,
and $k^\alpha_{\mu, \nu} = k^\alpha_{i,j}$ for $\mu = i ~ \text{(mod N)}$ and $\nu = j ~ \text{(mod N)}$.
The range of (integer) summation index $\xi$ depends on the existing pathways for $\I$-cycle.
Hereafter, the superscript $\I$ on $\mu$ shall be omitted for simplicity.

Following Ref. \cite{Koza1999}, we define
\begin{equation}
	P_j(\mu, t)\equiv  P(\mu,t) \delta^N_{\mu,j}
\end{equation}
where,
\begin{equation}
	\delta^N_{\mu,j} = 
	\begin{cases}
		1 \text{, if } j = \mu \text{ (mod $N$) }\\
		0 \text{, otherwise}
	\end{cases}
\end{equation}
Here, $j \in \{1, 2, \cdots, N\}$.
Multiplying  $\delta^N_{\mu, j}$ on both sides of Eq.\ref{eq:master} and using the equality $\delta_{\mu,j}^N = \delta_{\mu-\xi, j-\xi}^N$, we get
\begin{widetext}
	\begin{align} \label{eq:ME}
		\frac{\partial P_j(\mu, t)}{\partial t}= \sum_{\xi}  K_{j -\xi, j} P_{j-\xi}( \mu -\xi, t)  - K_{j, j-\xi} P_j(\mu, t). 
	\end{align}
\end{widetext}

\subsection{ Generating function }
We define a generating function to derive $V$ and $D$. 
The generating function for $\I$-cycle is defined by
\begin{equation} \label{eq:genQ}
	\mathcal{G}^{\mathcal{I}}_j (z,t) \equiv \sum_{\mu=-\infty}^{\infty} e^{z X^{\mathcal{I}}_\mu} P_j (\mu, t).
\end{equation}
where $X^{\mathcal{I}}_\mu$ denotes the generalized coordinate for $\I$-cycle at generalized state $\mu$.
Then Eq.(\ref{eq:ME}) and the equality $(X^\I_{\mu} - X^\I_{\mu-\xi}) P_{j-\xi}(\mu-\xi, t) = (X^\I_j- X^\I_{j-\xi})P_{j-\xi}(\mu-\xi, t) $ with $\delta_{\mu,j}^N = \delta_{\mu-\xi, j-\xi}^N$ lead to  
\begin{widetext}
	\begin{equation}
		\begin{aligned} \label{eq:MEq}
			\frac{\partial \mathcal{G}^{\mathcal{I}}_j(z, t)}{\partial t} 
			&=\sum_{\xi} \left( \sum_{\mu=-\infty}^{\infty} e^{z X^{\mathcal{I}}_\mu} K_{j -\xi, j} P_{j-\xi}( \mu -\xi, t) \right)
			- \sum_{\xi} K_{j, j-\xi} \mathcal{G}^{\mathcal{I}}_j(z, t) \\
			&= \sum_{\xi} e^{ z d^\mathcal{I}_{j- \xi, j} }  K_{j -\xi, j} \mathcal{G}^{\mathcal{I}}_{j-\xi}( z, t) 
			- \sum_{\xi} K_{j, j-\xi} \mathcal{G}^{\mathcal{I}}_j(z, t) \\
		\end{aligned}
	\end{equation}
\end{widetext}
where $d^\mathcal{I}_{\mu \nu} \equiv X^{\mathcal{I}}_\nu - X^{\mathcal{I}}_\mu$.
In general, different cycle has different $\{d^\mathcal{I}_{\mu, \nu}\}$.
For example, for the $\mathcal{F}$-cycle in Fig. \ref{fig_kinesin_6state}A,
\begin{equation}
	\begin{aligned}
		d^\mathcal{F}_{i, j} = 
		\begin{cases}
			1 \text{, for $i=6, j=1$} \\
			-1 \text{, for $i=1, j=6$} \\
			0 \text{, otherwise}, 
		\end{cases}
	\end{aligned}
\end{equation}
for the $\mathcal{B}$-cycle,
\begin{equation}
	\begin{aligned}
		d^\mathcal{B}_{i, j} = 
		\begin{cases}
			1 \text{, for $i=3, j=4$} \\
			-1 \text{, for $i=4, j=3$} \\
			0 \text{, otherwise}, 
		\end{cases}
	\end{aligned}
\end{equation}
and for the $\mathcal{X}$-cycle, 
\begin{equation}
	\begin{aligned}
		d^\mathcal{X}_{i, j} = 
		\begin{cases}
			1 \text{, for $i=2, j=5$} \\
			-1 \text{, for $i=5, j=2$} \\
			0 \text{, otherwise}.   
		\end{cases}
	\end{aligned}
\end{equation}

In fact, Eq. (\ref{eq:MEq}) can be expressed more succinctly as 
\begin{equation}\label{eq:MEqM}
	\begin{aligned}
		\partial_t \mathcal{G}^{\mathcal{I}}_j(z, t) = \sum_{i=1}^{N}\Gamma^\mathcal{I}_{i j} \mathcal{G}^{\mathcal{I}}_i(z,t)
	\end{aligned}
\end{equation}
where 
\begin{equation} \label{eq:Gamma} 
	\Gamma^\mathcal{I}_{i j}(z) = 
	\begin{cases}
		\sum_{\alpha} k^\alpha_{i j} e^{z d^{\mathcal{I},\alpha}_{i j}} \text{, if $i \neq j$}\\
		- \sum_{m=1(\neq i)}^N \sum_{\alpha} k^\alpha_{i m}\text{, if $i = j$}
	\end{cases}
\end{equation}
With $\alpha$, an index to discern the pathways, 
$\Gamma$ can be written in the form of $N \times N$ matrix.

\subsection{Generating function at the asymptotic limit}
Here we consider the asymptotic limit ($t \ra \infty$) in which $V$ and $D$ are well defined for an arbitrary chemical network model.
The general solution of Eq.(\ref{eq:MEqM}) can be written as \cite{Koza1999}
\begin{equation}
	\begin{aligned}
		\mathcal{G}^{\mathcal{I}}_j(z,t) = \sum_m T^\mathcal{I}_{m j}(z, t) e^{\lambda^\mathcal{I}_m(z) t}
	\end{aligned}
\end{equation}
where $\lambda^\mathcal{I}_m(z)$'s ($m=0,1,2,\ldots$, $N$) are the eigenvalues of $\Gamma^\mathcal{I}(z)$.
For a system in (unique) steady state, the eigenvalues satisfy $\lambda^\mathcal{I}_0(0) = 0$ and $\lambda^\mathcal{I}_m(0) < 0$ for $m\neq 0$.
Thus, at $t \rightarrow \infty$ and when $z \sim 0$,
\begin{equation}\label{eq:QApprox}
	\begin{aligned}
		\lim_{t \rightarrow \infty} \mathcal{G}^{\mathcal{I}}_j(z,t) \sim T^\mathcal{I}_{0 j}(z,t) e^{\lambda^\mathcal{I}_0(z) t}
	\end{aligned}
\end{equation}
Now, summed over the index $j$, Eq.(\ref{eq:genQ}) is led to
\begin{equation}
	\begin{aligned}
		\sum_{j=1}^N \mathcal{G}^{\mathcal{I}}_j(z,t) &= \sum_{j=1}^N \sum_{\mu=-\infty}^{\infty} e^{z X^\mathcal{I}_\mu} P_j(\mu, t) \\
		&= \sum_{\mu=-\infty}^{\infty} e^{z X^\mathcal{I}_\mu} P(\mu, t) \\
		&\equiv \mathcal{G}^\mathcal{I}(z,t). 
	\end{aligned}
\end{equation}
From Eq.(\ref{eq:QApprox}), at $t \rightarrow \infty$, we have
\begin{equation}
	\begin{aligned} \label{eq:Q}
		\mathcal{G}^{\mathcal{I}}(z, t) &\sim \sum_j {T^\mathcal{I}_{0 j}(z,t)} e^{\lambda^{\mathcal{I}}_0(z) t} = h^{\mathcal{I}}(z,t) e^{\lambda^{\mathcal{I}}_0(z) t}
	\end{aligned}
\end{equation}
where $h^{\mathcal{I}}(z,t) \equiv \sum_j T_{0 j}^\mathcal{I}(z,t)$.
Since $\mathcal{G}^{\mathcal{I}}(0,t) = 1$ and $\lambda^\mathcal{I}_0(z=0)=0$, 
$h^{\mathcal{I}}(0,t) \sim 1$ at $t \rightarrow \infty$. 
\subsection{ Velocity and Diffusion coefficient}
In this section, we first define the flux $J^\I$  and the diffusion coefficient $D^\I$ of $\I$-cycle using $X^\I(t)$ at $t \ra \infty$.
Then by using the asymptotic form of the generating function, we will get the relation between $J^\I$ and $D^\I$, and the lowest eigenvalue $\lambda_0^\I(z)$.

The mean value of the generalized coordinate $X^{\mathcal{I}}(t)$ can be obtained using  
\begin{equation}
	\begin{aligned}
		\bave{X^{\mathcal{I}}(t)} 
		&= \partial_z \mathcal{G}^{\mathcal{I}}(z,t) \vert_{z=0}\sim  (h^{\mathcal{I}})' 
		+ t ~ (\lambda^{\mathcal{I}}_0)'
	\end{aligned}
\end{equation}
where Eq.(\ref{eq:Q}) was used and the prime denotes a partial derivative with respect to $z$ at $z=0$.
The flux of $\I$-cycle is defined by
\begin{equation}
	\begin{aligned}
		J_{\mathcal{I}} &\equiv \lim_{t \rightarrow \infty} \frac{ \bave{X^{\mathcal{I}}(t) }}{t}=(\lambda^{\mathcal{I}}_0)'.
	\end{aligned}
\end{equation}
$J_\mathcal{X}$ multiplied by the step size $d_0$ corresponds to the velocity $V$ of motor 

Similarly, the diffusion coefficient $D_\I$ is obtained by considering the second moment of $X^\I$.
\begin{align}
	&\bave{ (X^{\mathcal{I}})^2 } = \partial_z^2 \mathcal{G}^{\mathcal{I}}(z,t)\nonumber\\
	&= (h^{\mathcal{I}})'' + 2 t (h^{\mathcal{I}})' (\lambda^{\mathcal{I}}_0)' + t (\lambda^{\mathcal{I}}_0)'' + ((\lambda^{\mathcal{I}}_0)')^2 t^2, 
\end{align}
which gives  
\begin{equation}
	\begin{aligned}
		D_\mathcal{I} =\lim_{t \rightarrow \infty} \frac{\bave{(X^\mathcal{I}(t))^2}  - \bave{ X^\mathcal{I} (t)}^2}{2 t} = \frac{(\lambda^{\mathcal{I}}_0)''}{2}.
	\end{aligned}
\end{equation}
Thus, the diffusion coefficient of motor is obtained: $D = d_0^2 D_\mathcal{X}$.
\\

\subsection{Characteristic polynomial}
To express the derivatives of $\lambda^{\mathcal{I}}_0$ in terms of rates $\{k_{ij}\}$, we use the characteristic polynomial of $\Gamma^\mathcal{I}(z)$ \cite{Koza1999},
\begin{align}
	\det( \lambda^\mathcal{I}_0 \mathbb{I} - \Gamma^\mathcal{I}(z) ) &= \sum_{n=0}^{N}  (\lambda^{\mathcal{I}}_0)^n C_n(z) = 0. 
	\label{eqn:det}
\end{align}
By differentiating both side of Eq.\ref{eqn:det} with respect to $z$ and setting $z=0$, we get
\begin{equation} \label{eq:cha_d1}
	\begin{aligned}
		C_0' + C_1 (\lambda^{\mathcal{I}}_0)' &= 0, \\
	\end{aligned}
\end{equation}
and 
\begin{equation} \label{eq:cha_d2}
	\begin{aligned}
		C_0'' + 2 C_1' ( \lambda^{\mathcal{I}}_0)' + C_1 (\lambda^{\mathcal{I}}_0)'' + 2 C_2 (( \lambda^{\mathcal{I}}_0)' )^2= 0.
	\end{aligned}
\end{equation}
From Eqs.(\ref{eq:cha_d1}) and (\ref{eq:cha_d2}), we get 
\begin{equation} \label{eq: V in C}
	J_{\mathcal{I}} = (\lambda^{\mathcal{I}}_0)' = - \frac{ C_0' }{ C_1 }
\end{equation}
\begin{widetext}
	\begin{equation} \label{eq: D in C}
		D_{\mathcal{I}} = \frac{(\lambda^{\mathcal{I}}_0)''}{2} = - \frac{ C_0'' + 2 C_1' (\lambda^{\mathcal{I}}_0)' + 2 C_2 (\lambda^{\mathcal{I}}_0)'}{2 C_1} = - \frac{C_0'' + 2 J_\mathcal{I}  + 2 C_2 J_\mathcal{I} }{ 2 C_1}
	\end{equation}
\end{widetext}
$C_n$'s and their derivatives, which depend on the choice of $X^{\mathcal{I}}$, can readily be found by differentiating the characteristic polynomial with respect to $\lambda^\mathcal{I}_0(z)$ with $\lambda^\I_0(0)=0$ \cite{Koza1999}.

\subsection{ Explicit expression of $J_\mathcal{F}$  }
The expression of reaction current in each subcycle $\mathcal{F}$ and $\mathcal{B}$ in terms of $\{k_{ij}\}$ can be obtained by considering the corresponding generating function $\mathcal{G}^{\I\in\{\F,\mathcal{B}}\}$
Here, we provide the expression of $J_\mathcal{F}$ in terms of rate constants $\{k_{ij}\}$ for the 6-state double-cycle kinetic network.
\begin{widetext}
	\begin{equation} \label{eq:JF}
		\begin{aligned}
			J_\mathcal{F}&=J_\mathcal{F}^+-J_\mathcal{F}^-\nonumber\\
			&=\bigg(k_{12} (k_{25} k_{32} k_{43} + k_{23} k_{34} k_{45} + k_{25} (k_{32} + k_{34}) k_{45}) k_{56} k_{61}
			-k_{16} k_{21} (k_{34} k_{45} k_{52}
			+ k_{32} (k_{43} + k_{45}) k_{52} + k_{32} k_{43} k_{54}) k_{65}\bigg) \bigg/
			\\
			&\bigg(((k_{25} k_{32} k_{43} + k_{23} k_{34} k_{45} + k_{25} (k_{32} + k_{34}) k_{45}) k_{56} 
			+ k_{21} (k_{34} k_{45} (k_{52} + k_{56}) 
			\\& + k_{32} (k_{45} (k_{52} + k_{56}) + k_{43} (k_{52} + k_{54} + k_{56})))) k_{61} 
			+ k_{21} (k_{34} k_{45} k_{52} + k_{32} (k_{43} + k_{45}) k_{52} + k_{32} k_{43} k_{54}) k_{65} 
			\\&+k_{16} ((k_{25} k_{32} k_{43} + k_{23} k_{34} k_{45} 
			+ k_{25} (k_{32} + k_{34}) k_{45}) k_{56} + (k_{32} k_{43} k_{52} + k_{32} k_{45} k_{52} 
			+ k_{34} k_{45} k_{52} + k_{32} k_{43} k_{54} 
			\\&+ 
			k_{25} (k_{34} k_{45} + (k_{34} + k_{43}) k_{54} + k_{32} (k_{43} + k_{45} + k_{54})) + 
			k_{23} ((k_{43} + k_{45}) k_{52} + k_{43} k_{54} + k_{34} (k_{45} + k_{52} + k_{54}))) k_{65} 
			\\&+
			k_{21} (k_{34} k_{45} (k_{52} + k_{56}) + k_{43} k_{54} k_{65} + k_{34} (k_{45} + k_{54}) k_{65} + 
			k_{32} (k_{54} k_{65} + k_{45} (k_{52} + k_{56} + k_{65})
			\\& + 
			k_{43} (k_{52} + k_{54} + k_{56} + k_{65})))) + 
			k_{12} ((k_{34} k_{45} (k_{52} + k_{56}) + 
			k_{32} (k_{45} (k_{52} + k_{56}) + 
			k_{43} (k_{52} + k_{54} + k_{56}))) k_{61} 
			\\&+ (k_{34} k_{45} k_{52} + 
			k_{32} (k_{43} + k_{45}) k_{52} + k_{32} k_{43} k_{54}) k_{65} + 
			k_{23} (k_{45} (k_{52} + k_{56}) k_{61} 
			\\&+ k_{43} (k_{52} + k_{54} + k_{56}) k_{61} + 
			k_{45} k_{52} k_{65} + k_{43} (k_{52} + k_{54}) k_{65} 
			+ 
			k_{34} ((k_{52} + k_{54} + k_{56}) k_{61} + (k_{52} + k_{54}) k_{65}
			\\& + 
			k_{45} (k_{56} + k_{61} + k_{65}))) + 
			k_{25} (k_{43} k_{54} (k_{61} + k_{65}) + 
			k_{34} (k_{54} (k_{61} + k_{65}) + k_{45} (k_{56} + k_{61} + k_{65})) 
			\\&+ 
			k_{32} (k_{54} (k_{61} + k_{65}) + k_{43} (k_{56} + k_{61} + k_{65}) + 
			k_{45} (k_{56} + k_{61} + k_{65}))))\bigg)
		\end{aligned}
	\end{equation}
\end{widetext}
Similarly, $J_\mathcal{B}$ and $D_\mathcal{\mathcal{X}}$ can also be expressed in terms of $\{k_{ij}\}$. 
\\
\section{ Analysis of other types of kinesins }
\subsection{ Kinesin-1 mutant (Kin6AA) }
Single molecule motility data digitized from Ref. \cite{Clancy:2011:NSMB} was fitted to 6-state network model (Figs. \ref{fig_kinesin_6state}A, \ref{fig_exp_Kin6AA}) by using the same method employed for the analysis of kinesin-1 data ({\bf Materials and Methods }).
However, 4 additional initial conditions for $k_{25}$ ($\{300, 3000, 30000, 3000000\}$), thus total 245 initial conditions, were explored.  
The rate constants estimated from this procedure are provided in Table \ref{table}.

\subsection{ Kinesin-2 (KIF17, KIF3AB) }
Single molecule motility data digitized from Ref. \cite{Milic:2017:PNAS} was again fitted to the 6-state double-cycle kinetic model (Fig. \ref{fig_kinesin_6state}A, \ref{fig_exp_KIF17}, \ref{fig_exp_KIF3AB}) following the identical procedure employed in the analysis of kinesin-1 data ({\bf Materials and Methods}).
However, two additional initial conditions for $k_{25}$ ($\{30000, 3000000\}$) were explored, which results in total 147 initial conditions.
The rate constants are shown in Table \ref{table}.

\section{Myosin-V}
Here we summarize the multi-cyclic model for myosin-V \cite{Bierbaum:2011:BPJ} which consists of  ATP-dependent chemomechanical  forward cycle $\mathcal{F}$, dissipative cycle $\mathcal{E}$, and ratcheting cycle (ATP independent stepping cycle)  $\mathcal{M}$ (Fig. \ref{fig_Qa_other_motor_proteins}A).
We first provide the explanation of how $V$ and $D$ of myosin-V are calculated.
Next, the affinity and heat production ($\dot{Q}$) are expressed in terms of a set of rates $\{k_{ij}\}$.
Finally, $\mathcal{Q}$ shall be calculated using $V$, $D$, and $\dot{Q}$. 

\subsection{Calculation of $V$ and $D$}
The $\mathcal{M}$-cycle consisting of a single state (Fig. \ref{fig_Qa_other_motor_proteins}A) prevents the application of Eq.(\ref{eq:Gamma}).
To circumvent this difficulty, the model with additional state ($5'$) is considered (Fig. \ref{fig_myosin_model_aug}).
The $(5')$-state is chemically equivalent to the state (5), but describes motor in different position on actins, such that $X(5) = X_0$ and $X(5') = X_0 \pm d_0$ where $d_0 = 36$ nm for myosin-V.
In this new network, the rate constants $\kappa_{ij}$'s are
\bea{
	\kappa_{2, 5} &= \kappa_{2, 5'} = \frac{k_{25}}{2} \\
	\kappa_{6, 5} &= \kappa_{6, 5'} = \frac{k_{65}}{2} \\
	\kappa_{5, 2} &= \kappa_{5', 2} = k_{52} \\
	\kappa^f_{5,5'} &= \kappa^f_{5', 5} = k_{55, f} \\
	\kappa^b_{5,5'} &= \kappa^b_{5', 5} = k_{55, b} \\
}
\\
where the subscripts $f$ and $b$ denote the forward and backward motion, respectively.
Other rate constants satisfy $\kappa_{ij} = k_{ij}$.
This modification can be justified by considering stochastic movement of myosin-V on the chemical network \cite{Gillespie:1977}:
$\kappa_{i,5}, \kappa_{i,5'}$ are set to $k_{i5}/2$, such that the outgoing fluxes from the states $i=$ (2), (6) to the state (5) remain identical in the both networks depicted in Fig. \ref{fig_Qa_other_motor_proteins}A and Fig.\ref{fig_myosin_model_aug}. 
Next, we set $\kappa_{5,i} = \kappa_{5', i}$ to keep the inward fluxes toward (6), (2) identical for the two networks.
Finally, $\kappa^{f}_{5, 5'} = \kappa^{f}_{5', 5} = k_{55, f}$ and $\kappa^{b}_{5, 5'} = \kappa^{b}_{5', 5} = k_{55, b}$. 
These modification of rate constants enable us to describe transitions within the $\M$-cycle.

Now, the elements of distance matrix scaled by $d_0$ are
\bea{
	d^\mathcal{X}_{3,4} & = 1,\\
	d^\mathcal{X}_{4,3} & = -1, \\
	d^{\mathcal{X},f}_{5, 5'} &= 1,\\
	d^{\mathcal{X},f}_{5', 5} &= 1,\\
	d^{\mathcal{X},b}_{5, 5'} &= -1, \\
	d^{\mathcal{X},b}_{5', 5} &= -1. \\
}
Other elements ($d^\mathcal{X}_{i,j}$) are all zero.
Thus, $\Gamma_{i,j}^\mathcal{X}$ is written as (with $(7) \equiv (5')$)
\\
\beaw{
	\left(
	\begin{smallmatrix} \label{eq:Gamma_mV}
		-{\kappa_{12}}-{\kappa_{14}} & {\kappa_{12}} & 0 & {\kappa_{14}} & 0 & 0 & 0 \\
		{\kappa_{21}} & -{\kappa_{21}}-{\kappa_{23}}-{\kappa_{25}}-{\kappa_{26}}-{\kappa_{27}} & {\kappa_{23}} & 0 & e^z {\kappa_{25}} & {\kappa_{26}} & {\kappa_{27}} \\
		0 & {\kappa_{32}} & -{\kappa_{32}}-{\kappa_{34}} & {\kappa_{34}} & 0 & 0 & 0 \\
		{\kappa_{41}} & 0 & {\kappa_{43}} & -{\kappa_{41}}-{\kappa_{43}} & 0 & 0 & 0 \\
		0 & e^{-z} {\kappa_{52}} & 0 & 0 & -{\kappa_{52}}-{\kappa_{56}}-{\kappa_{57,b}}-{\kappa_{57,f}} & {\kappa_{56}} & {\kappa_{57,b}}+{\kappa_{57,f}} \\
		0 & {\kappa_{62}} & 0 & 0 & {\kappa_{65}} & -{\kappa_{62}}-{\kappa_{65}}-{\kappa_{67}} & {\kappa_{67}} \\
		0 & {\kappa_{72}} & 0 & 0 & {\kappa_{75,b}}+{\kappa_{75,f}} & {\kappa_{76}} & -{\kappa_{72}}-{\kappa_{75,b}}-{\kappa_{75,f}}-{\kappa_{76}}
	\end{smallmatrix}
	\right)
}

Now, the travel velocity $V$ and the diffusion coefficient $D$ of myosin-V are readily acquired  by using Eqs.(\ref{eq: V in C}, \ref{eq: D in C}, and \ref{eq:Gamma_mV}).
The rate constants used in the calculation are summarized in Table. \ref{table_mV}.
%
%
\subsection{Affinities and heat production}
The affinities of individual cycles are 

\bea{
	\A_\F &= k_B T \log \pave{ \frac{k_{12} k_{23} k_{34} k_{41} }{ k_{21} k_{32} k_{43} k_{41} } }, 
	\\
	\A_\E &= k_B T \log \pave{\frac{k_{25} k_{56} k_{62} } {k_{52} k_{65} k_{26} } }, \\
	\A_\M &= k_B T \log \pave{\frac{k_{55,f} } {k_{55,b} } }. \\
}
\\
Only the  following rate constants depend on the load ($f$): 
\bea{
	k_{34} &= k_{34}^o e^{- \theta d_m f / k_B T } \\
	k_{43} &= k_{43}^o e^{(1-\theta) d_m f / k_B T }\\
	k_{56} &= k_{56}^o \frac{ 1 + e^{-\chi d_m f_c / k_B T } }{ 1 + e^{ \chi d_m (f-f_c) / k_B T } }\\
	k_{52} &= k_{52}^o \frac{ 1 + e^{-\chi d_m f_c / k_B T } }{ 1 + e^{ \chi d_m (f-f_c) / k_B T } }\\
	k_{55,b} &= \frac{ D'}{k_B T } \frac{ f d_m - U}{d_m^2} \frac{1}{1 - e^{ \pave{ U - f d_m} /k_B T }} \\
	k_{55,f} &= k_{55,b} e^{- f d_m / k_B T}
}
where 
\bea{
	\theta &= 0.65\\
	\chi &= 4\\
	f_c & = 1.6 \text{ pN } \\
	U &= \text{20 }k_B T \\
	D' & = 4.7 \times 10^{-4} \mu m/s^2
} 
as described in Ref. \cite{Bierbaum:2011:BPJ}.
Thus, the affinities can be written as
\bea{
	\A_\F &= k_B T   \log \pave{ \frac{k_{12}^o k_{23}^o k_{34}^o k_{41}^o }{ k_{21}^o k_{32}^o k_{43}^o k_{41}^o}} - f d_0 \\
	\A_\E &= \A_{\E, f=0}\\
	\A_\M &= - f d_0. 
}
The relation $\A_\M = - f d_0$ results from the fact that $\M$-cycle is ATP-independent and activated by the load.
Thus, the heat production rate of the system is 
\bea{
	\dot{Q} &= J_\F \A_F + J_\E \A_E + J_\M  \A_M \\
}
$J_\F, J_\E$, and $J_\M$ can be calculated by using Eqs. (\ref{eq: V in C}) and (\ref{eq:Gamma_mV}).
Finally, $\mathcal{Q}$ for myosin-V is given by 
\bea{
	\mathcal{Q}_{\text{Myosin-V} } &= \dot{Q}\frac{ 2 D } {V^2}  \\
}
where $D = D_\mathcal{X} d_0^2$ and $V = J_\mathcal{X} d_0$.
\ref{table_dynein}

\section{ Dynein}
$(N=7)$-unicyclic model is considered based on the model of cytoplasmic dimeric dynein studied in Ref. \cite{Sarlah:2014:BPJ}.
Only the major forward pathway, where the transitions between the states are denoted by solid black lines in Fig. 5A of Ref. \cite{Sarlah:2014:BPJ}, is considered.
The values of rate constants obtained from Ref. \cite{Sarlah:2014:BPJ} are summarized in Table \ref{table_dynein}.
To describe the force-dependence of power-stroke, we model the rate constant for forward and reverse strokes ($k_{+PS}$ and $k_{-PS}$) as follows.
\begin{equation} \label{eq:f_dynein}
	\begin{aligned}
		k_{+PS} &= k_{+PS, f=0} e^{-\theta \frac{f d_0}{k_B T} }\\
		k_{-PS} &= k_{-PS, f=0} e^{(1-\theta) \frac{f d_0}{k_B T} }\\
	\end{aligned}
\end{equation} 
where $\theta = 0.3$ is selected based on the previous studies \cite{Singh:2005:PNAS,Wagoner:2016:JPCB}.
In the original literature \cite{Sarlah:2014:BPJ}, all the rate constants depend on both elastic energy originated from the interaction between two monomer units of dynein, and $f$. 
Although this approach will better describe the details of dynein dynamics, it is not possible to calculate elastic energy without explicit simulation of the motion of dyneins which are modeled as elastic materials \cite{Sarlah:2014:BPJ}.
Thus, for simplicity, we assumes only $k_{\pm PS}$ changes significantly by $f$.
Again, $V$ and  $D$ were calculated using Eqs. \ref{eq: V in C}, \ref{eq: D in C}.
\subsection{ Affinity and heat production }
The affinity for unicyclic model is written as \cite{Qian2005_BiophyChem,Seifert:2005:PRL,Hwang2017JPCL}
\bea{
	\mathcal{A} &= k_B T \log \prod_{i=1}^N \frac{ k_{i,i+1}} {k_{i+1, i}}
	\\& = -\Delta \mu_{\text{hyd}} - f d_0
}
where $d_0 = 8.2$ nm. 
The second term, describing force-dependence, is originated from the use of Eq. \ref{eq:f_dynein}.
Finally, $\mathcal{Q}$ is
\bea{
	\mathcal{Q} &= \frac{2 D}{V d_0} \mathcal{A}. 
}
\\

\section{F$_1$-ATPase}
Here, we summarize the unicyclic model developed for F$_1$-ATPase in Ref. \cite{Gerritsma:2010}.
The model is $(N=2)$ unicyclic model (Fig. \ref{fig_Qa_other_motor_proteins}C) where 3 cycles in chemical state space correspond to a single rotation in real space (angle changes by 90$\degree$ upon transition from the state (1) to state (2) whereas transitions from the state (2) to $(1)'$ induce 30$\degree$ rotation (Fig. \ref{fig_Qa_other_motor_proteins}C).
The model is valid when the torque applied to F$_1$-ATPase is small enough ($\tau \lesssim 30$ pN$\cdot$nm) that the mechanical cycle is tightly coupled to the chemical reaction \cite{Gerritsma:2010}.
The dependences of rate constants on the torque are 
\bea{
	k_{12}(\tau, \zeta) & = k^{\text{bi}}_{12} (\tau, \zeta) \times  [\text{ATP}] \\
	&= \frac{1}{ e^{a_{k_{12}}(\tau)} + \zeta e^{b_{k_{12}}(\tau)} }\times [\text{ATP}] \\
	k_{21'}(\tau, \zeta) & = \frac{1}{ e^{a_{k_{21'}}(\tau)} + \zeta e^{b_{k_{21'}}(\tau)} } \\
	k_{1'2}(\tau, \zeta) & = k_{1'2}^{\text{bi}}(\tau, \zeta) \times [\text{ADP}][\text{Pi}] \\
	&= \frac{1}{ e^{a_{k_{1'2}}(\tau)} + \zeta e^{b_{k_{1'2}}(\tau)} } \times [\text{ADP}][\text{Pi}]\\
	k_{21}(\tau, \zeta) &= \frac{ k_{12}(\zeta, \tau) u_2(\zeta, \tau) }{k_{21}(\zeta, \tau)} e^{(\Delta \mu_{\text{hyd}}^0 + \frac{2 \pi}{3} \tau)/k_B T}
}
\\
where $\Delta \mu_{\text{hyd}}^0 =-12.5$ $k_B T \approx -50$ pN$\cdot$nm, $\zeta$ is the friction coefficient (for example, if the $\gamma$-shaft of F$_1$-ATPase is attached to a bead of radius $r$, $\zeta = 2 \pi \eta r^3 ( 4 + 3 \sin^2 \pi/6 )$ \cite{Gerritsma:2010} with the water viscosity $\eta = 1$ cP $= 10^{-9}$ pN$\times$s$\times$ nm$^{-2}$. In our calculation, $r=40$ nm as in Ref. \cite{Gerritsma:2010}), and $a_i(\tau), b_i(\tau)$ are polynomial function of $\tau$ defined in Ref. \cite{Gerritsma:2010}. The expressions of $a_i(\tau)$, $b_i(\tau)$ and the coefficients of the polynomials are given in Table. \ref{table_f1}.

\subsection{ $V$, $D$, affinities, and heat production }
For (N=2)-unicyclic model, the speed of rotation $V$, diffusion coefficient $D$, and affinity $\mathcal{A}$ are \cite{Derrida1983_JSP,Koza1999,Fisher1999_PNAS,Qian07ARPC,Qian2005_BiophyChem,Seifert2012RPP,Hwang2017JPCL}
\begin{align} \label{eq:V,D,A_f1}
	V &= d_R \frac{k_{1,2} k_{2,1'} - k_{2,1} k_{1', 2}}{k_{1,2} + k_{2,1'} + k_{2,1} + k_{1', 2}} \equiv d_R J,\nonumber\\
	D&=\frac{d_R^2}{2}\left[\frac{k_{1,2}k_{2,1'}}{k_{2,1}k_{1', 2}}+1-2\left(\frac{k_{1,2}k_{2,1'}}{k_{2,1}k_{1', 2}}-1\right)^2\frac{k_{2,1}k_{1', 2}}{\sigma^2}\right]\nonumber\\
	&\times\frac{k_{2,1}k_{1', 2}}{\sigma}, \nonumber\\
	\A &= k_B T \log \pave{ \frac{ k_{1,2} k_{2,1'} }{ k_{2,1} k_{1', 2} } }\nonumber\\
	&=\pave{ -\Delta \mu_{\text{hyd}}^0 +  k_B T \log \pave{ \frac{ [ATP] }{[ADP][Pi]} }   } - \frac{2 \pi }{3} \tau \nonumber\\
	&= -\Delta \mu_{\text{hyd}} - W,
\end{align}
where $d_R = \frac{2 \pi}{3} $ is the radian distance that motor travels upon ATP hydrolysis, $\sigma=k_{12} + k_{2 1'} + k_{21} + k_{1'2}$, and $W \equiv  \frac{2 \pi}{3} \tau$ denotes the work done by the motor.
Here, $\tau>0$ implies the motor performs work against the hindering load.
Thus, $\mathcal{Q}$ is given by 
\bea{
	\mathcal{Q} = \frac{ 2 D  }{  V d_R } \mathcal{A}  
}
.

\section{ Unicyclic kinetic model for kinesin-1}
To analyze the kinesin-1 data, we also considered $(N=4)$-unicyclic model (Fig. \ref{fig_kinesin_uni}A) which was used in our previous study \cite{Hwang2017JPCL}.
Briefly, the model consists of four forward rates $\{u_1, u_2, u_3, u_4\}$ and four backward rates $\{w_1, w_2, w_3, w_4\}$.
Only $u_1 (= k^{bi} [\text{ATP}])$ depends on [ATP].
Barometric dependence of the rates on forces are assumed: $u_n = u_n^o e^{-f d_0 \theta_n^+ / k_B T}$ and $w_n = w_n^o e^{f d_0 \theta_n^- / k_B T}$ with $\sum_{n=1}^N\left( \theta_n^+ + \theta_n^- \right) = 1$ \cite{Fisher99PNAS,Fisher01PNAS}. 
$V$, $D$, $A$, and a set of rate constants used in the calculation of $\mathcal{Q}$ are provided in Table. \ref{table_kinesin_uni} \cite{Hwang2017JPCL}.

\section{ Unicyclic kinetic model for myosin-V}
For myosin-V, we also considered the ($N=2$) unicyclic model from Ref. \cite{Kolomeisky:2003:BPJ}. 
Briefly, the model consists of two forward rates $\{u_1, u_2\}$ and two backward rates $\{w_1, w_2\}$.
Only $u_1 (= k [\text{ATP}])$ and $w_2 (= k'[\text{ATP}]^\alpha)$ depend on [ATP].
Here, $\alpha = 1/2$. 
Different choice of $\alpha$ introduces only minor difference in the results as argued in \cite{Kolomeisky:2003:BPJ}.
Barometric dependences of the rates on forces are assumed again: $u_n = u_n^o e^	3{-f d_0 \theta_n^+ / k_B T}$ and $w_n = w_n^o e^{f d_0 \theta_n^- / k_B T}$ with $\sum_{n=1}^N\left( \theta_n^+ + \theta_n^- \right) = 1$ \cite{Fisher99PNAS,Fisher01PNAS}. 
The parameters used in the calculation are available in Eqs. (12), (13) in Ref. \cite{Kolomeisky:2003:BPJ} and summarized in Table. \ref{table_mV_uni}.
Identical expressions for $V$, $D$, and $\mathcal{A}$ from Eq. \ref{eq:V,D,A_f1} were used for the calculation except for $W = f d_0$. 

\section{The lower bound of $\mathcal{Q}$ for unicyclic model }
The analytic expression for the lower bound of the uncertainty measure $\mathcal{Q}$ is available for unicyclic models \cite{barato2015PRL}.
For $(N)$-state unicyclic model, the lower bound of $\mathcal{Q}$ is 
\begin{equation} \label{eq:Qb}
	\mathcal{Q}_b = \frac{ \mathcal{A} }{N} \coth \pave{ \frac{\mathcal{A}} { 2 N k_B T} } \geq 2 k_B T.
\end{equation}
The $\mathcal{Q}_b$ and the $\Delta \mathcal{Q} \equiv \mathcal{Q} - \mathcal{Q}_b$ of the motors as a function of $f$ and [ATP] are calculated in Figs. \ref{fig_kinesin_uni}D (kinesin-1), \ref{fig_f1_various}D (F$_1$-ATPase), and \ref{fig_myosin_uni}D (myosin-V).

%


\begin{figure*}[ht]
	\centering
	\includegraphics[scale=0.62]{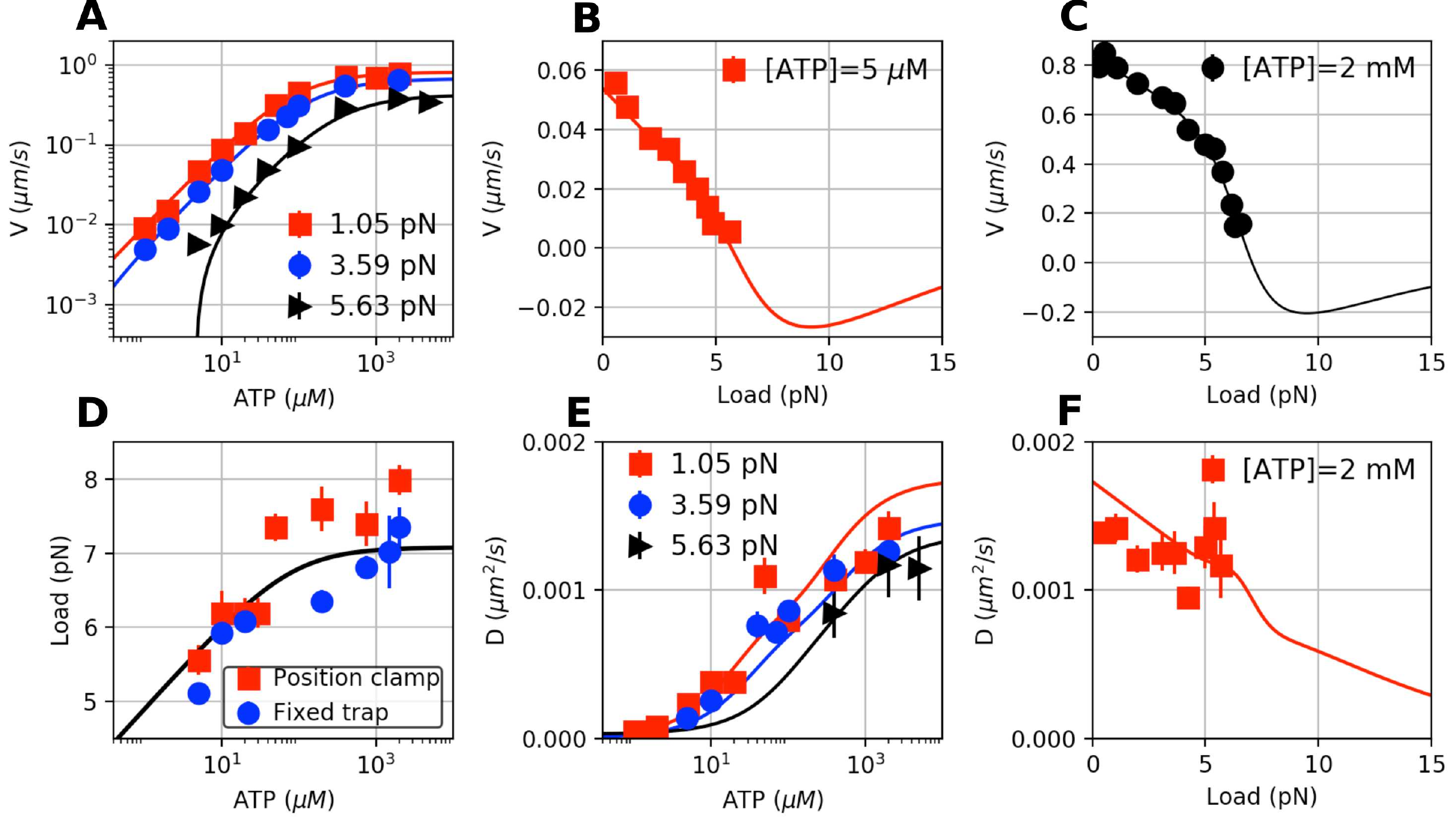}
	\caption{Analysis of experimental data of kinesin-1, digitized from Ref. \cite{Visscher99Nature}, using the 6-state model \cite{Liepelt07PRL}.
		The solid lines are the fits to the data
		{\bf A.} $V$ vs [ATP] at $f=$1.05 pN (red square), 3.59 pN (blue circle), and 5.63 pN (black triangle).
		{\bf B.} $V$ vs $f$ at [ATP] = 5 $\mu$M.
		{\bf C.} $V$ vs $f$ at [ATP] = 2 mM.
		{\bf D.} Stall force as a function of [ATP], measured by `Position clamp' (red square) or `Fixed trap' (blue circle) methods.
		{\bf E.} $D$ vs [ATP] at $f=$1.05 pN (red square), 3.59 pN (blue circle), and 5.63 pN (black triangle).
		$D$ was estimated from $r=2D/Vd_0$. 
		{\bf F.} $D$ vs $f$ at [ATP] = 2 mM.
	}
	\label{fig_exp_data_fit}
\end{figure*}

\begin{figure*}[ht]
	\centering
	\includegraphics[scale=0.85]{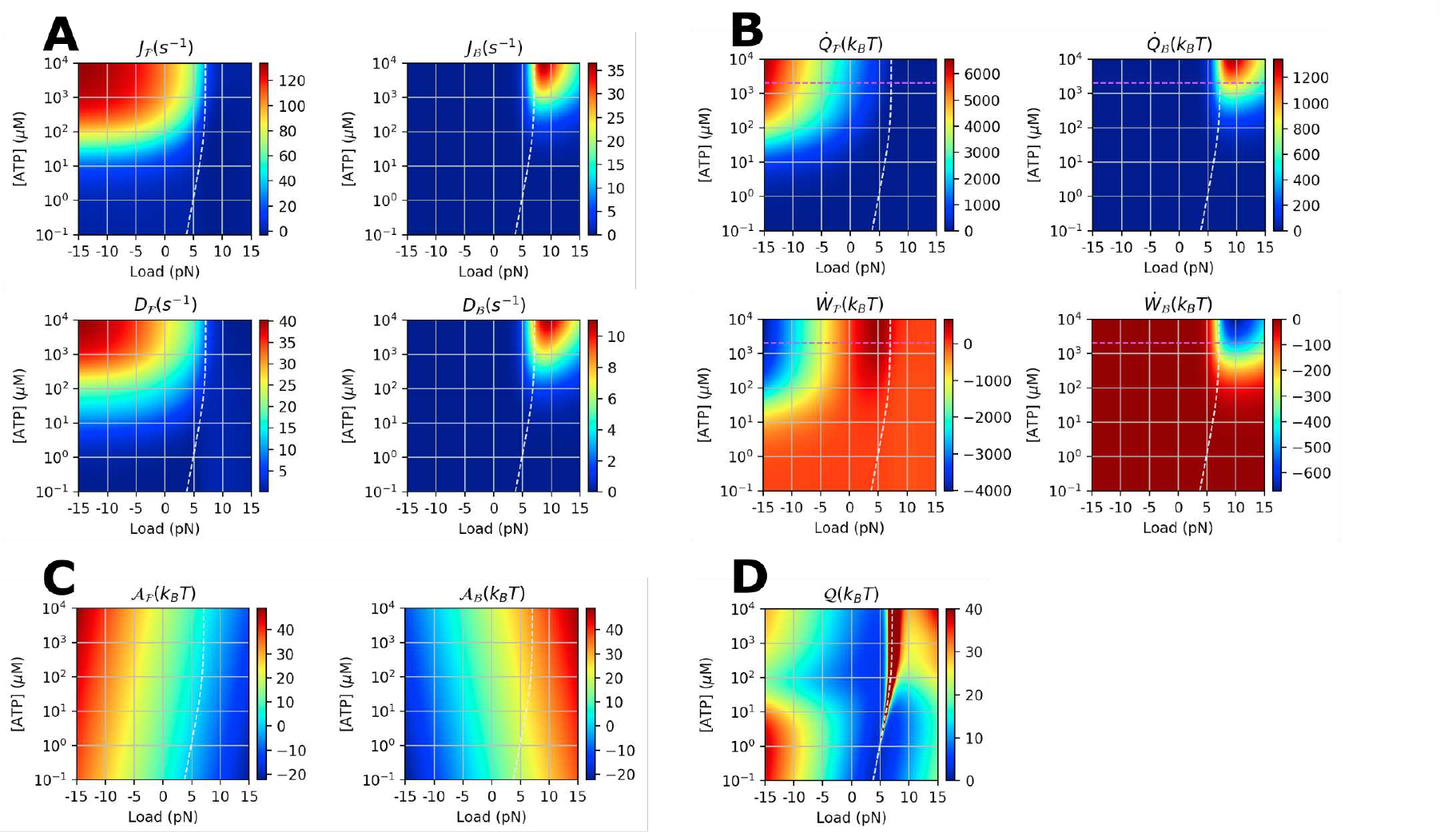}
	\caption{ 
		Various physical properties of kinesin-1 calculated using the 6-state network model (Fig. \ref{fig_kinesin_6state}A) at varying $f$ and [ATP].
		{\bf A}. Transport properties (flux $J$ and $D$).
		{\bf B}. Heat $\dot{Q}$ and work production $\dot{W}$.
		{\bf C}. Thermodynamic affinity $\mathcal{A}$.
		{\bf D.} $\mathcal{Q} (f, \text{[ATP]})$.
		The white dashed lines indicate the stall condition.
	}
	\label{fig_QWVD6_supple}
\end{figure*} 

\begin{figure*}[ht]
	\centering
	\includegraphics[scale=0.9]{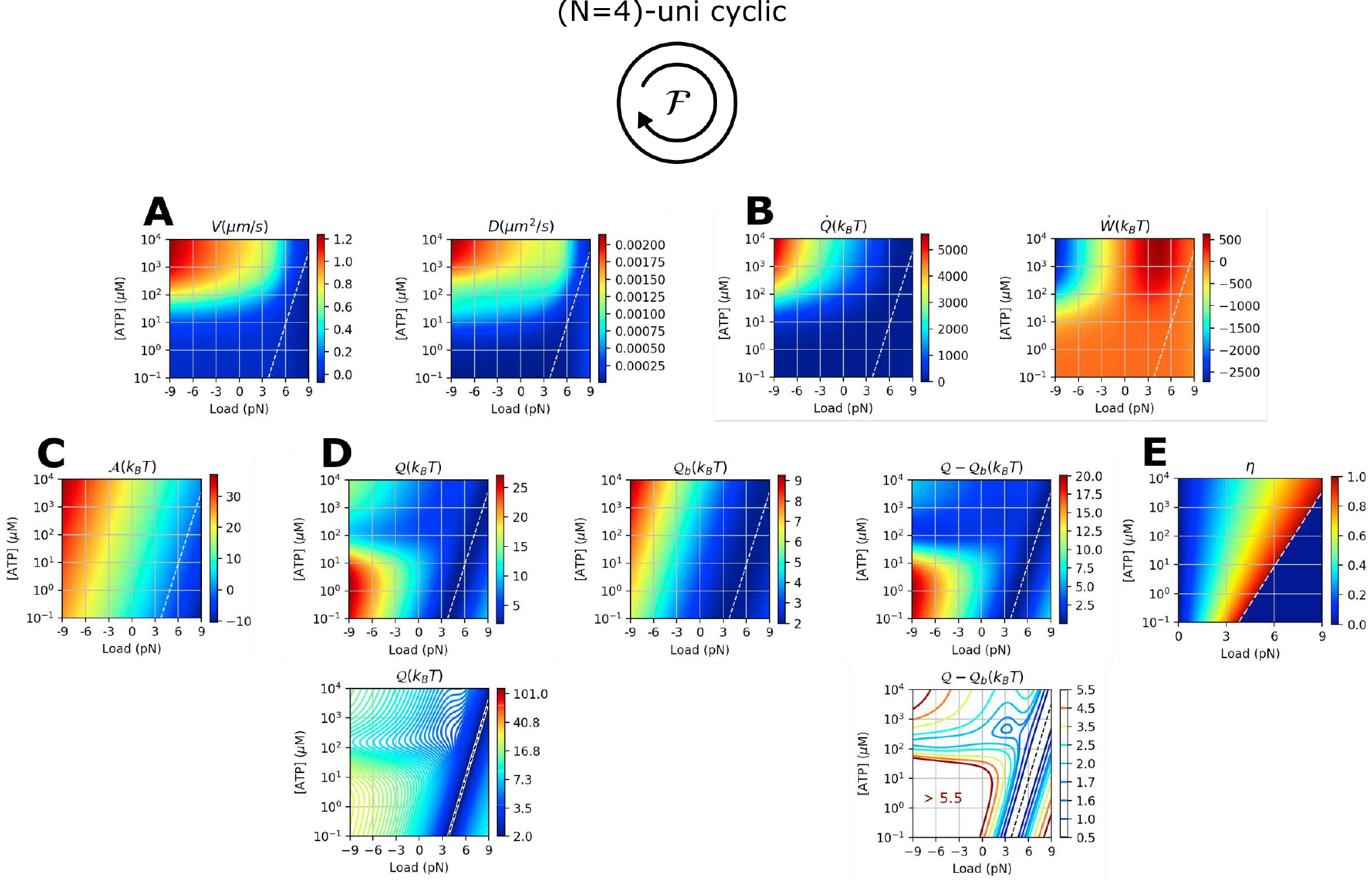}
	\caption{
		Various physical properties of kinesin-1 calculated using (N=4)-unicycle model \cite{Fisher01PNAS,Hwang2017JPCL} at varying $f$ and [ATP]. 
		We used the same model parameters from Ref. \cite{Hwang2017JPCL}. 
		{\bf A}. Transport properties ($V$, $D$).  
		{\bf B}. Heat $\dot{Q}$ and work production $\dot{W}$.
		{\bf C}. Thermodynamic affinity $\mathcal{A}$. 
		{\bf D}. $\mathcal{Q}$, $\mathcal{Q}_b$ (Eq.\ref{eq:Qb}), and their difference $\Delta \mathcal{Q}=\mathcal{Q}-\mathcal{Q}_b$.
		For clarity, identical data of $\mathcal{Q}$ and $\Delta \mathcal{Q}$ are shown again using contour plots.
		{\bf E}. Power efficiency $\eta \equiv \dot{W}/(\dot{W}+\dot{Q})$ ($\eta=0$ for $f > f_\text{stall}$). 
		The white dashed lines indicate the stall condition.
	}
	\label{fig_kinesin_uni}
\end{figure*} 

\begin{figure*}[ht]
	\centering
	\includegraphics[scale=0.4]{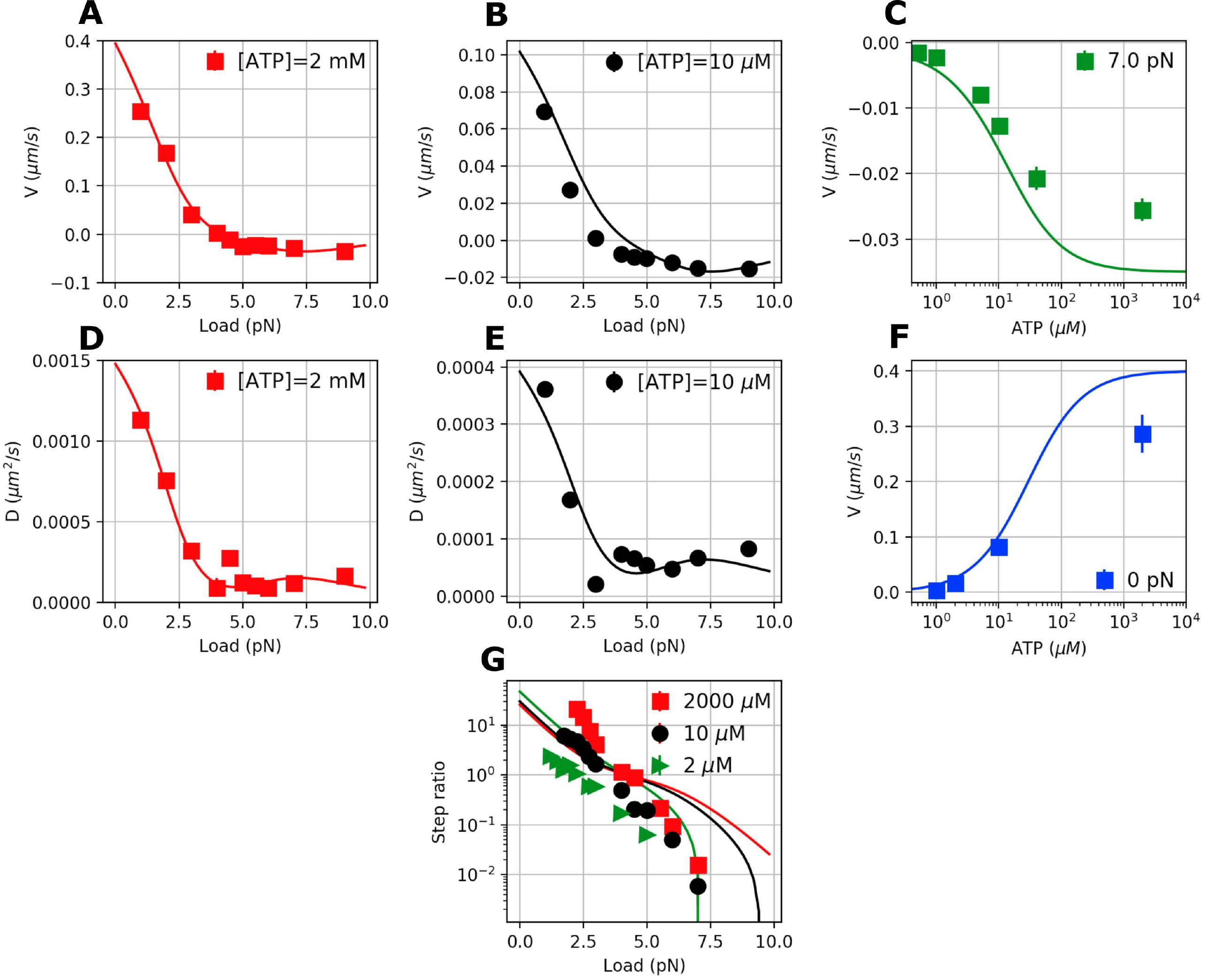}
	\caption{Motility data of Kin6AA, a mutant made of kinesin-1 to which six additional amino-acids are inserted in the neck-linker domains \cite{Clancy:2011:NSMB}, and the theoretical fits made using the 6-state double-cycle kinetic network model (Fig. \ref{fig_kinesin_6state}A).
	}
	\label{fig_exp_Kin6AA}
\end{figure*} 

\begin{figure*}[ht]
	\centering
	\includegraphics[scale=1]{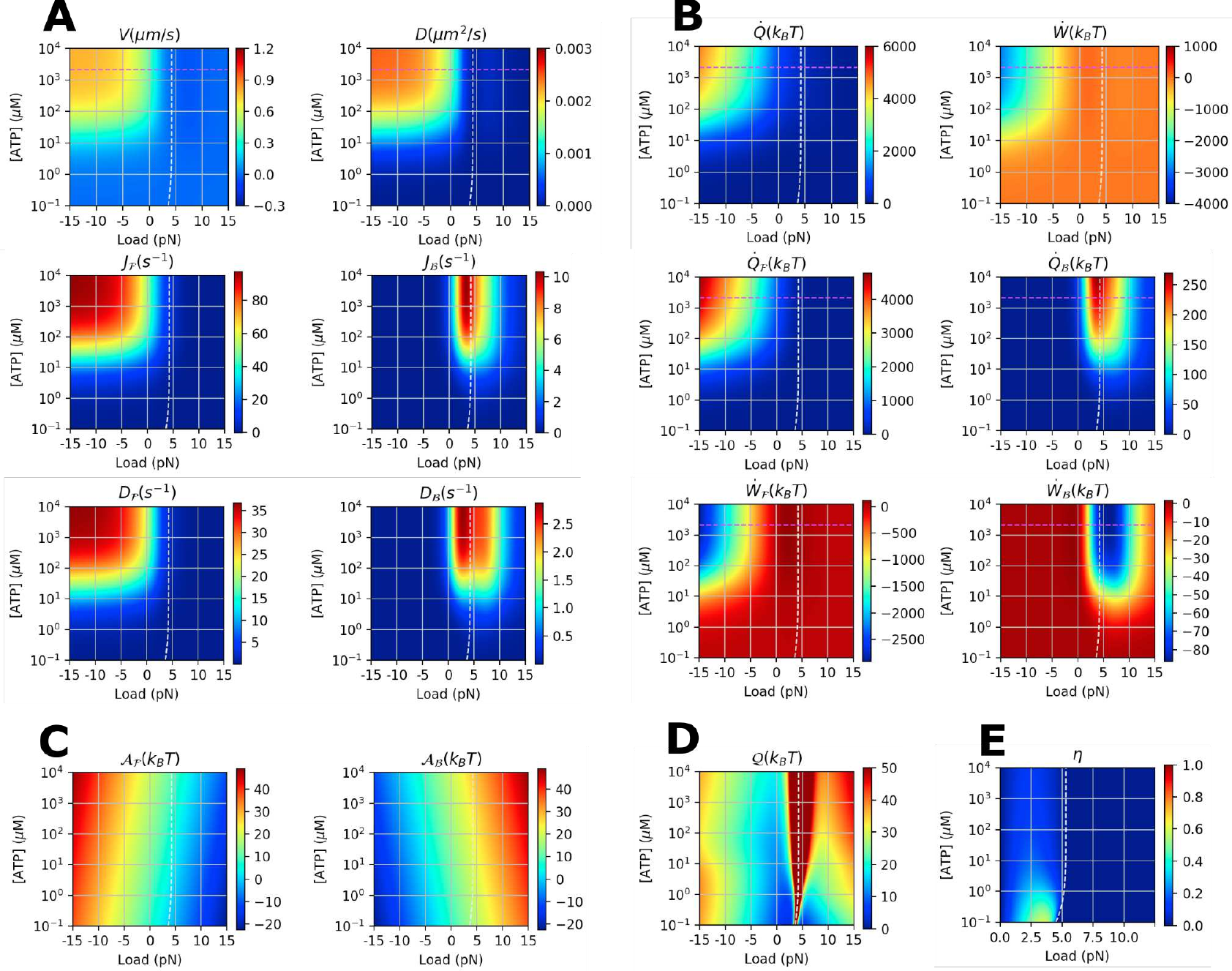}
	\caption{
		Various physical properties of kin6AA calculated using 6-state network model (Fig. \ref{fig_kinesin_6state}A) at varying $f$ and [ATP].
		{\bf A}. Transport properties ($J$, $D$).
		{\bf B}. Heat $\dot{Q}$ and work production $\dot{W}$.
		{\bf C}. Thermodynamic affinity $\mathcal{A}$.
		{\bf D.} $\mathcal{Q} (f, \text{[ATP]})$.
		{\bf E.} Power efficiency $\eta \equiv \dot{W}/(\dot{W}+\dot{Q})$ ($\eta=0$ for $f > f_\text{stall}$).
		The white dashed lines indicate the stall condition.
	}
	\label{fig_Kin6AA_various}
\end{figure*} 

\begin{figure*}[ht]
	\centering
	\includegraphics[scale=0.4]{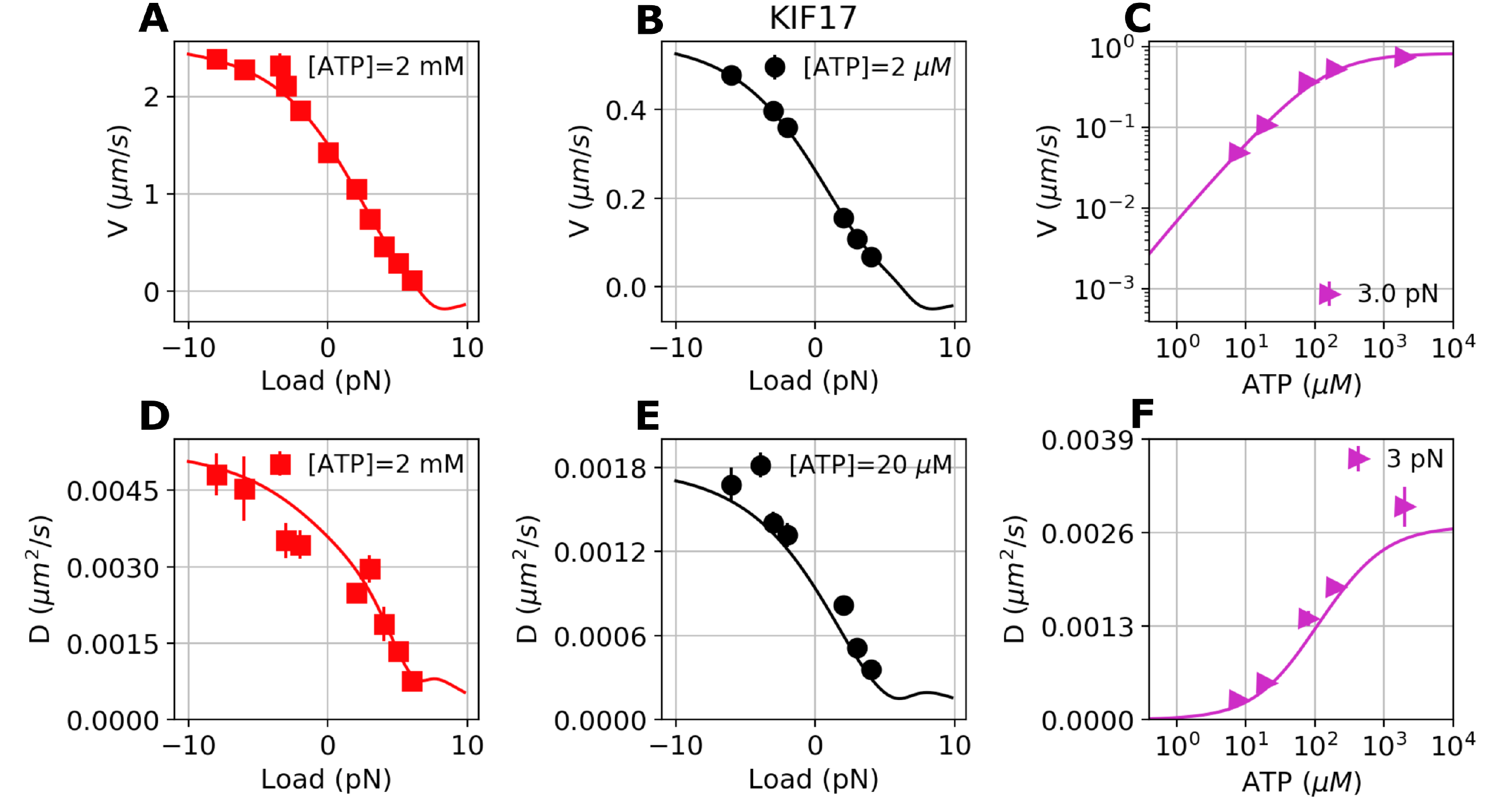}
	\caption{
		Motility data of homodimeric kinesin-2 (KIF17) \cite{Milic:2017:PNAS} and their theoretical fits (solid lines).
	}
	\label{fig_exp_KIF17}
\end{figure*} 

\begin{figure*}[ht]
	\centering
	\includegraphics[scale=1]{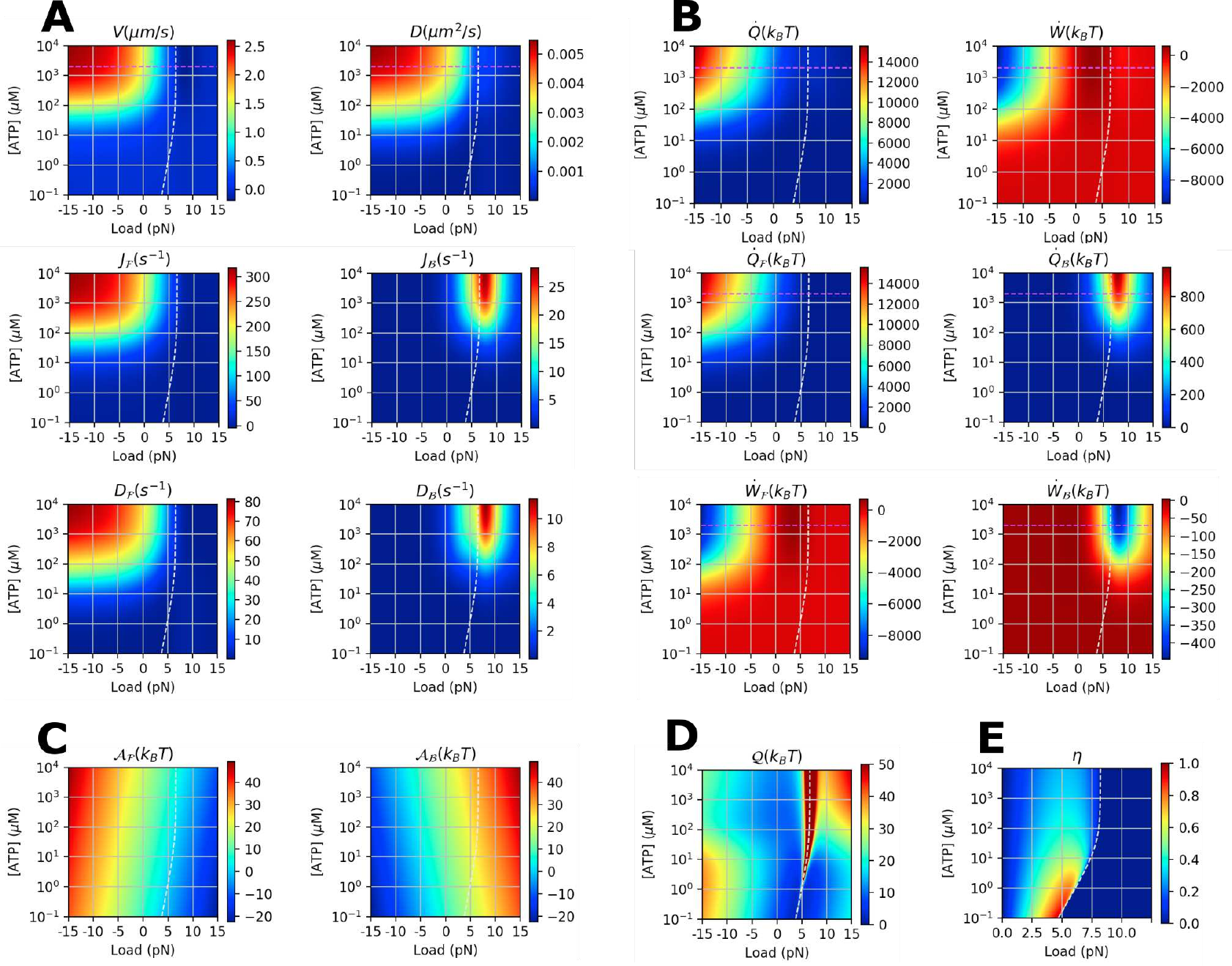}
	\caption{Various physical properties of KIF17 calculated using 6-state network model (Fig. \ref{fig_kinesin_6state}A) at varying $f$ and [ATP].
		{\bf A}. Transport properties ($J$, $D$).
		{\bf B}. Heat $\dot{Q}$ and work production $\dot{W}$.
		{\bf C}. Thermodynamic affinity $\mathcal{A}$.
		{\bf D.} $\mathcal{Q} (f, \text{[ATP]})$.
		{\bf E.} Power efficiency $\eta \equiv \dot{W}/(\dot{W}+\dot{Q})$ ($\eta=0$ for $f > f_\text{stall}$).
		The white dashed lines indicate the stall condition.
	}
	\label{fig_KIF17_various}
\end{figure*} 

\begin{figure*}[ht]
	\centering
	\includegraphics[scale=0.4]{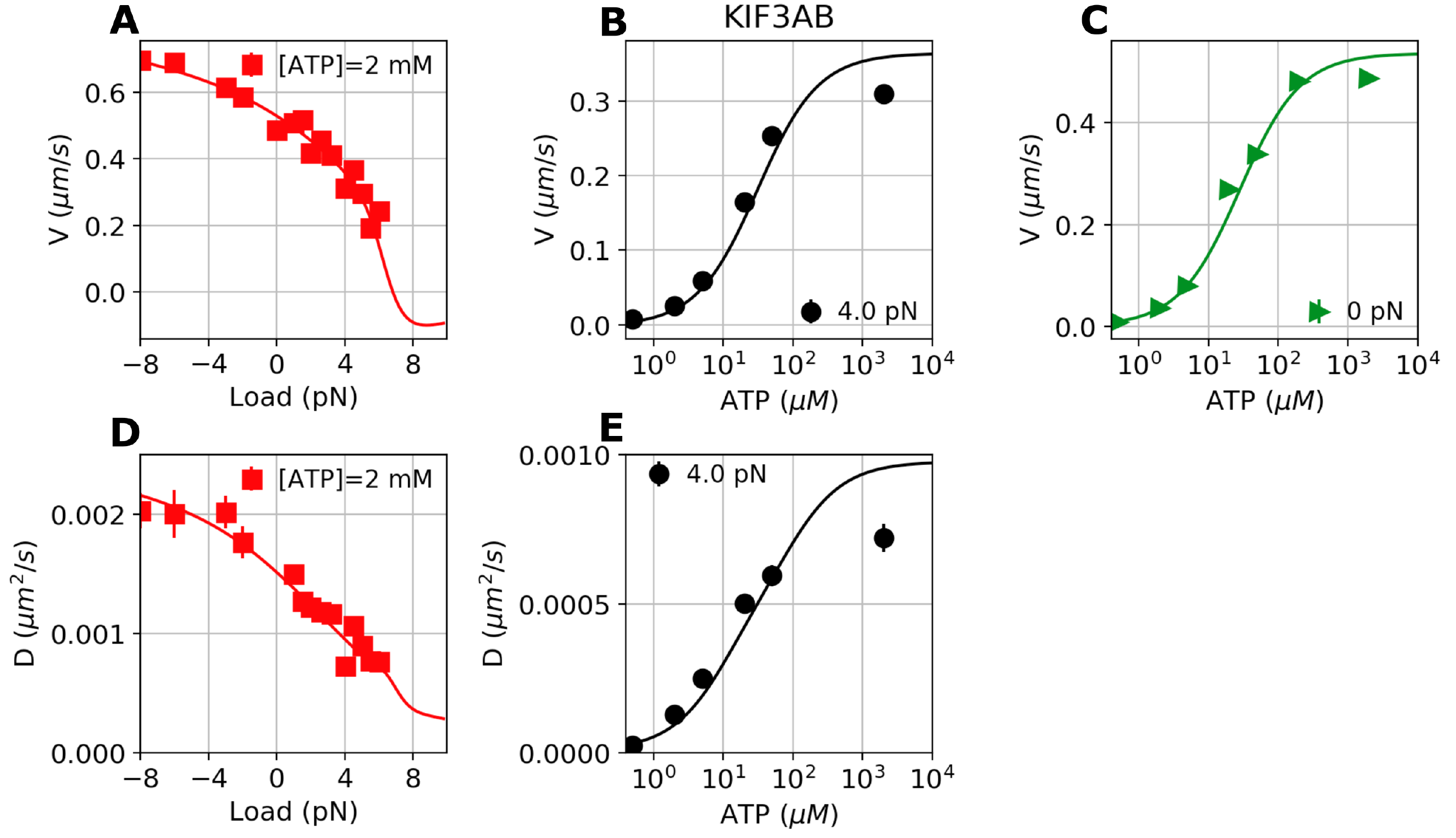}
	\caption{Motility data of heterotrimeric kinesin-2 (KIF3AB) \cite{Milic:2017:PNAS} and their theoretical fits (solid lines).
	}
	\label{fig_exp_KIF3AB}
\end{figure*} 

\begin{figure*}[ht]
	\centering
	\includegraphics[scale=1]{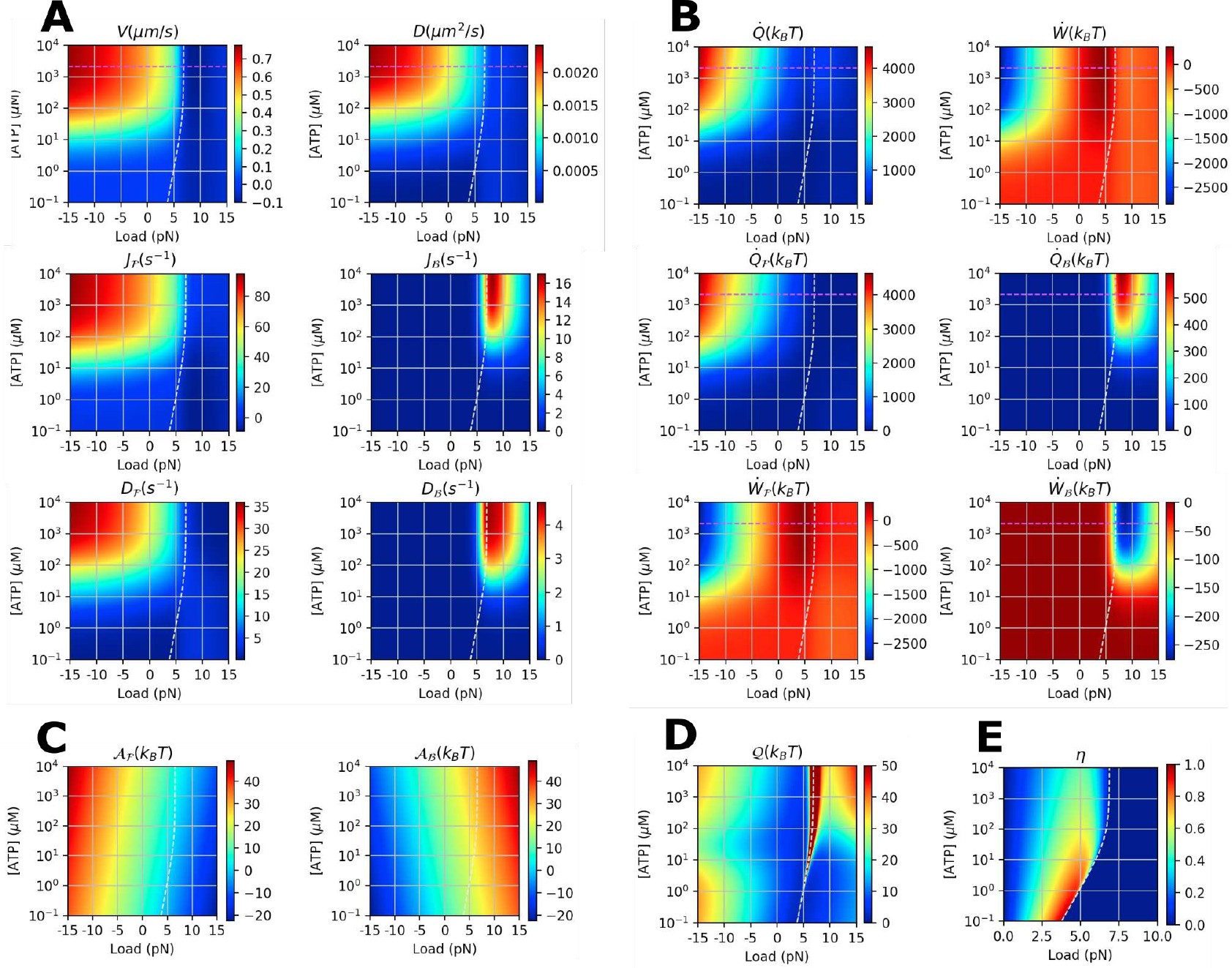}
	\caption{Various physical properties of KIF3AB calculated using the 6-state network model (Fig. \ref{fig_kinesin_6state}A) at varying $f$ and [ATP].
		{\bf A}. Transport properties ($J$, $D$).
		{\bf B}. Heat $\dot{Q}$ and work production $\dot{W}$.
		{\bf C}. Thermodynamic affinity $\mathcal{A}$.
		{\bf D.} $\mathcal{Q} (f, \text{[ATP]})$.
		{\bf E.} Power efficiency $\eta \equiv \dot{W}/(\dot{W}+\dot{Q})$ ($\eta=0$ for $f > f_\text{stall}$).
		The white dashed lines indicate the stall condition.
	}
	\label{fig_KIF3AB_various}
\end{figure*}

\begin{figure*}[ht]
	\centering
	\includegraphics[scale=0.9]{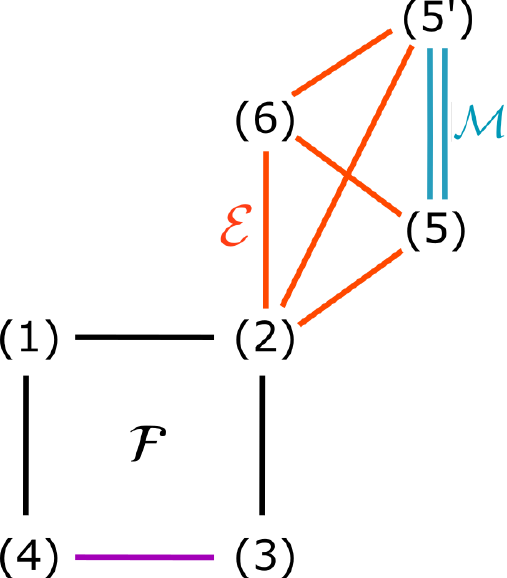}
	\caption{
		Augmented kinetic network model of myosin-V. Each line represents a reversible kinetics.
	}
	\label{fig_myosin_model_aug}
\end{figure*} 

\begin{figure*}[ht]
	\centering
	\includegraphics[scale=0.7]{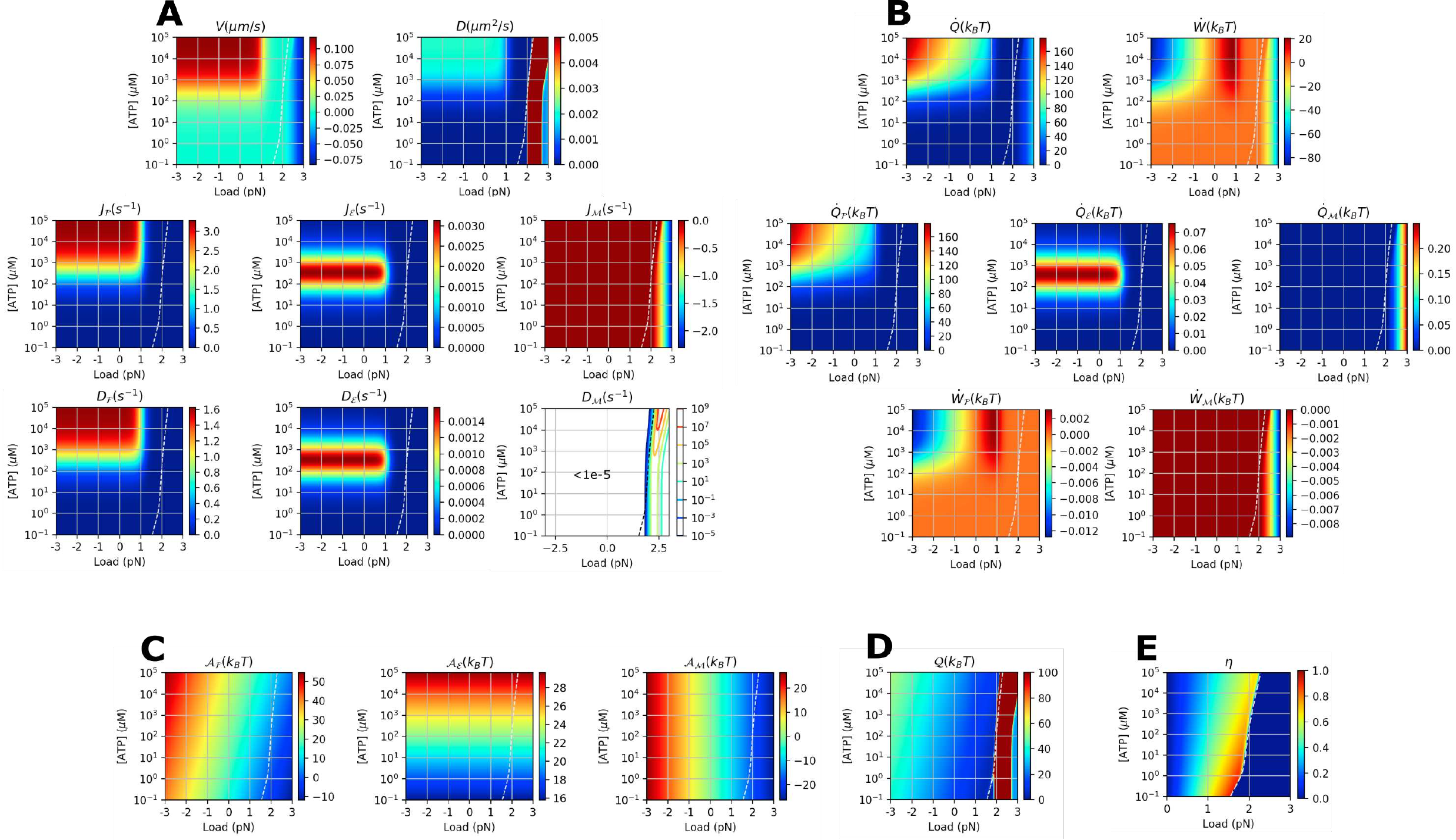}
	\caption{
		Various physical properties of myosin-V calculated using the multi-cyclic model \cite{Bierbaum:2011:BPJ} at varying $f$ and [ATP]. 
		[ADP] = 70 $\mu$M and [Pi] = 1 mM condition was used for the calculation.
		{\bf A}. Transport properties ($J$, $D$).
		{\bf B}. Heat $\dot{Q}$ and work production $\dot{W}$.
		{\bf C}. Thermodynamic affinity $\mathcal{A}$.
		{\bf D.} $\mathcal{Q} (f, \text{[ATP]})$.
		{\bf E.} Power efficiency $\eta \equiv \dot{W}/(\dot{W}+\dot{Q})$ ($\eta=0$ for $f > f_\text{stall}$).
		The white dashed lines indicate the stall condition.
	}
	\label{fig_myosin_various_70_1000}
\end{figure*} 

\begin{figure*}[ht]
	\centering
	\includegraphics[scale=0.7]{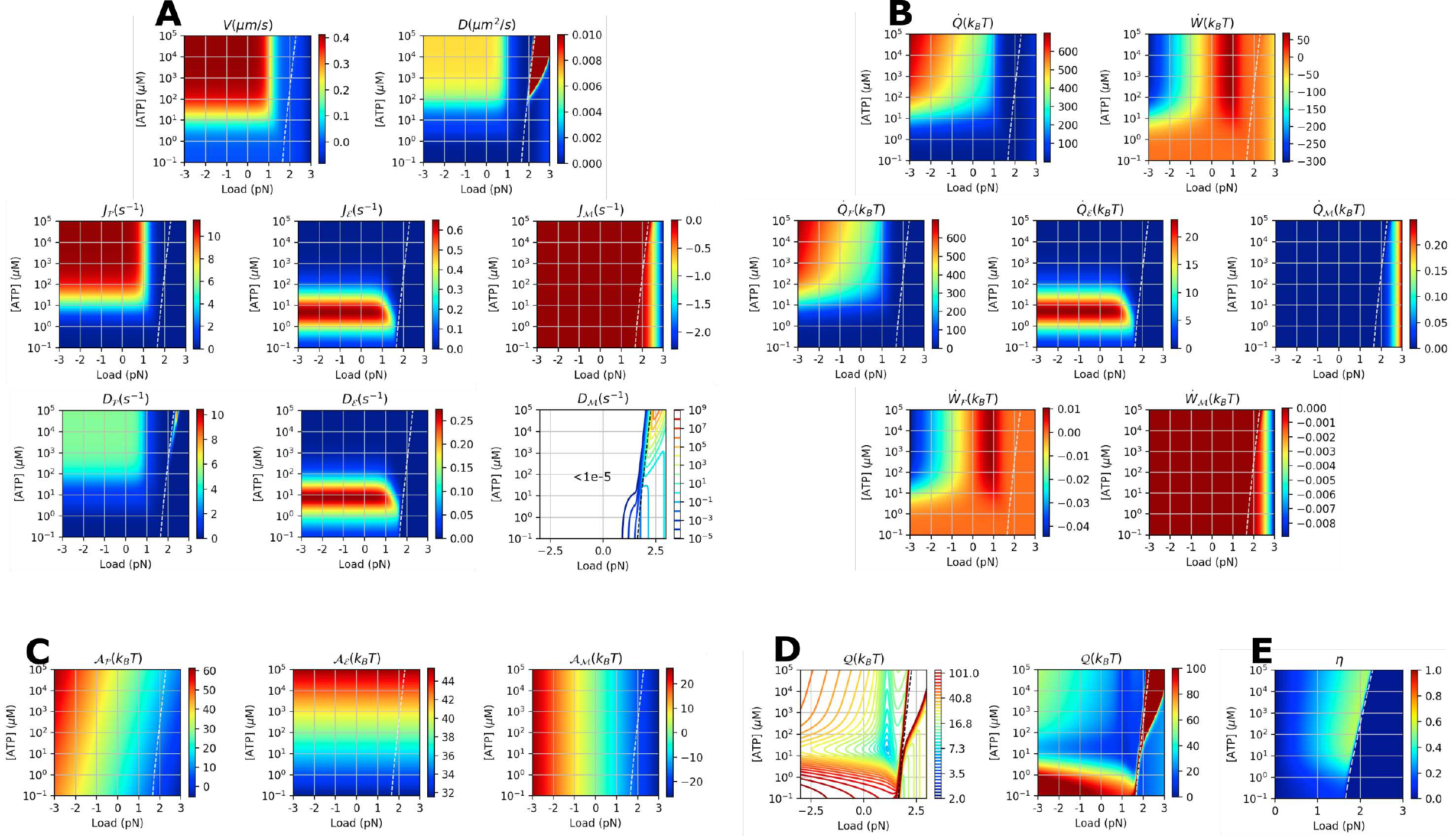}
	\caption{
		Various physical properties of myosin-V calculated using the multi-cyclic model \cite{Bierbaum:2011:BPJ} at varying $f$ and [ATP].
		[ADP] = 0.1 $\mu$M and [Pi] = 0.1 $\mu$M condition was used for the calculation.
		{\bf A}. Transport properties ($J$, $D$).
		{\bf B}. Heat $\dot{Q}$ and work production $\dot{W}$.
		{\bf C}. Thermodynamic affinity $\mathcal{A}$.
		{\bf D.} $\mathcal{Q} (f, \text{[ATP]})$.
		{\bf E.} Power efficiency $\eta \equiv \dot{W}/(\dot{W}+\dot{Q})$ ($\eta=0$ for $f > f_\text{stall}$).
		The white dashed lines indicate the stall condition.
	}
	\label{fig_myosin_various}
\end{figure*} 

\begin{figure*}[ht]
	\centering
	\includegraphics[scale=0.83]{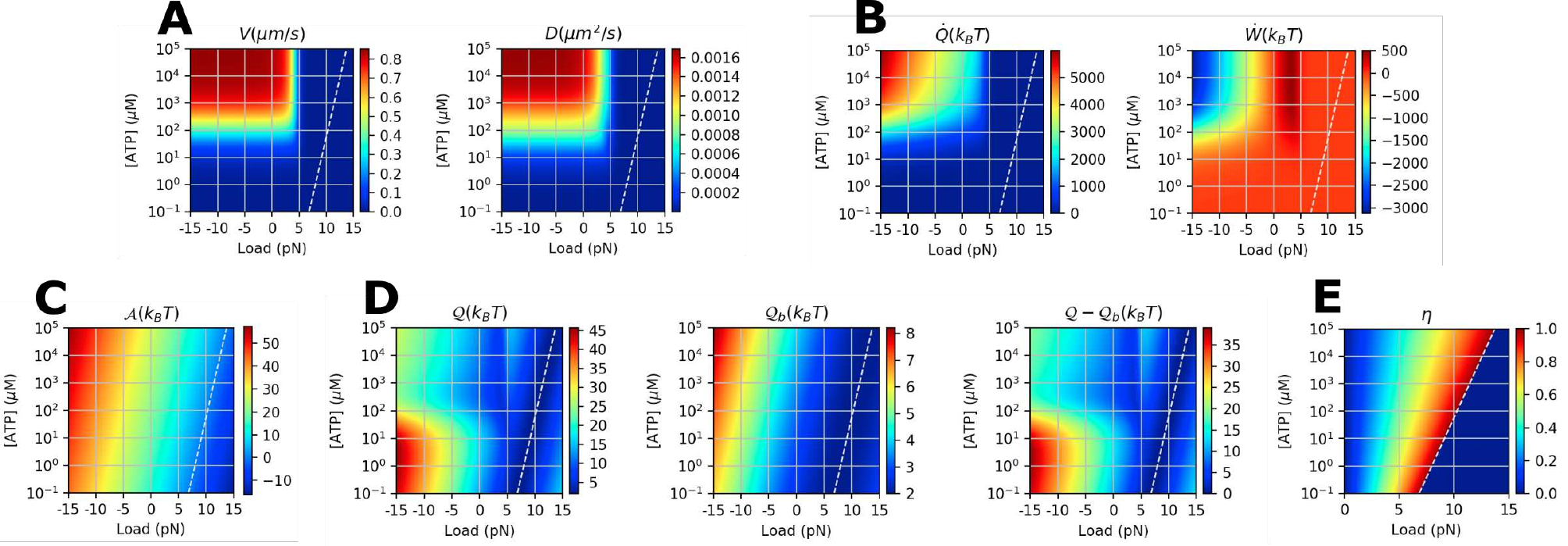}
	\caption{
		Various physical properties of dynein monomer calculated using ($N=7$)-unicyclic model \cite{Sarlah:2014:BPJ} at varying $f$ and [ATP].
		[ADP] = 70 $\mu$M and [Pi] = 1 mM condition was used for the calculation.
		{\bf A}. Transport properties ($V$, $D$).
		{\bf B}. Heat $\dot{Q}$ and work production $\dot{W}$.
		{\bf C}. Thermodynamic affinity $\mathcal{A}$.
		{\bf D.} $\mathcal{Q} (f, \text{[ATP]})$, $\mathcal{Q}_b(f,\text{[ATP]})$, and $\Delta \mathcal{Q}(f,[\text{ATP}]$.
		{\bf E.} Power efficiency $\eta \equiv \dot{W}/(\dot{W}+\dot{Q})$ ($\eta=0$ for $f > f_\text{stall}$).
		The white dashed lines indicate the stall condition.
	}
	\label{fig_dynein_various}
\end{figure*} 

\begin{figure*}[ht]
	\centering
	\includegraphics[scale=0.83]{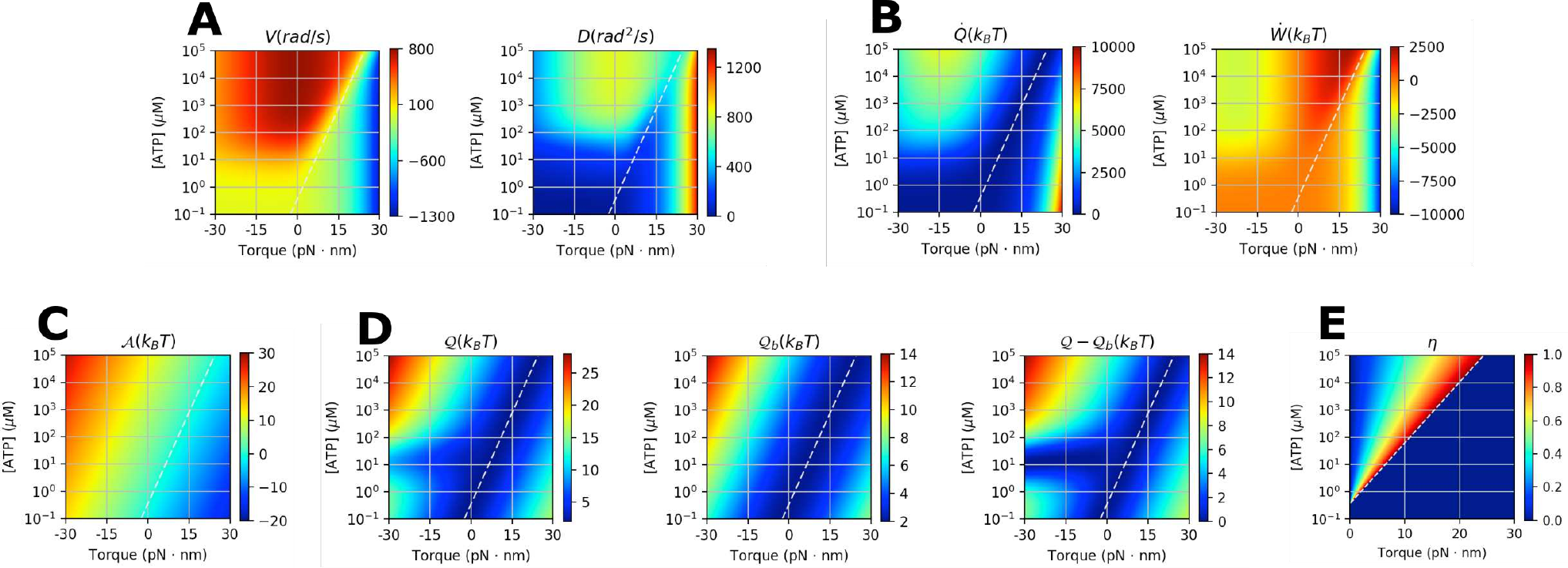}
	\caption{
		Various physical properties of F$_1$-ATPase calculated using ($N=2$)-unicyclic model (Fig. \ref{fig_Qa_other_motor_proteins}C) at varying $f$ and [ATP].
		[ADP] = 70 $\mu$M and [Pi] = 1 mM condition was used for the calculation.
		{\bf A}. Transport properties ($V$, $D$).
		{\bf B}. Heat $\dot{Q}$ and work production $\dot{W}$.
		{\bf C}. Thermodynamic affinity $\mathcal{A}$.
		{\bf D.} $\mathcal{Q}$, $\mathcal{Q}_b$ (Eq.\ref{eq:Qb}), and their difference $\Delta \mathcal{Q}=\mathcal{Q}-\mathcal{Q}_b$.
		{\bf E.} Power efficiency $\eta \equiv \dot{W}/(\dot{W}+\dot{Q})$ ($\eta=0$ for $f > f_\text{stall}$).
		The white dashed lines indicate the stall condition.
	}
	\label{fig_f1_various}
\end{figure*}

\begin{figure*}[ht]
	\centering
	\includegraphics[scale=0.8]{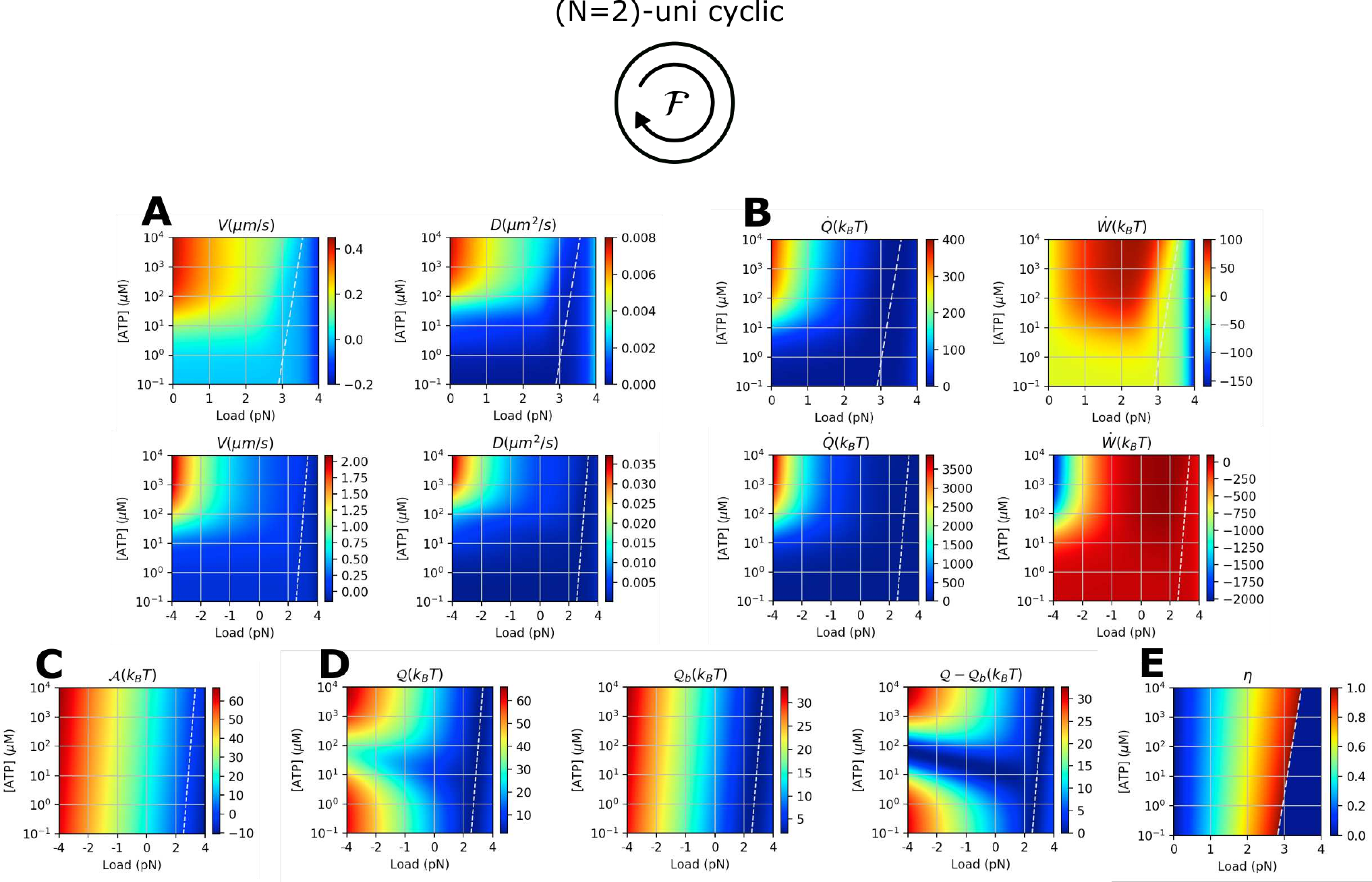}
	\caption{
		Various physical properties of myosin-V calculated using ($N=2$)-unicyclic model \cite{Kolomeisky:2003:BPJ} at varying $f$ and [ATP]. 
		{\bf A}. Transport properties ($V$, $D$). Same data are shown twice over the different range of $f$ for clarity.
		{\bf B}. Heat $\dot{Q}$ and work production $\dot{W}$. Same data are shown twice over the different range of $f$ for clarity.
		{\bf C}. Thermodynamic affinity $\mathcal{A}$.
		{\bf D.} $\mathcal{Q}$, $\mathcal{Q}_b$ (Eq.\ref{eq:Qb}), and their difference $\Delta \mathcal{Q}=\mathcal{Q}-\mathcal{Q}_b$.
		{\bf E.} Power efficiency $\eta \equiv \dot{W}/(\dot{W}+\dot{Q})$ ($\eta=0$ for $f > f_\text{stall}$).
		The white dashed lines indicate the stall condition.
	}
	\label{fig_myosin_uni}
\end{figure*} 


\clearpage

\begin{table*}[h!]
	\caption{Optimal $f$ and [ATP] that locally minimize $\mathcal{Q}$.
		For unicyclic models of kinesin-1, myosin-V, and F$_1$-ATPase, also shown are $f$ and [ATP] that  minimizes $\Delta \mathcal{Q} = \mathcal{Q}- \mathcal{Q}_b$, where $Q_b$ is a stronger lower bound for unicyclic kinetic schemes \cite{barato2015PRL} (Eq.\ref{eq:Qb}).
	}
	\footnotesize
	\begin{tabular}{|| c | c|c | c| c | c | c | c || }
		\midrule
		\hline
		&Kinesin-1 & Kinesin-1 & KIF17 &Myosin-V  & Myosin-V  & Dynein & F$_1$-ATPase 
		\\&(multi-cycle) & (unicycle) & (multi-cycle) &(multi-cycle) & (unicycle)   & (unicycle) & (unicyclic)
		\\& & & &[ADP]=[P$_i$]=0.1 $\mu$M & & &
		\\ \hline $f$ (pN)				 			 & 4.1 		   &	3.2 \footnote{\label{dQ}Condition for local minimization of  $\Delta \mathcal{Q} = \mathcal{Q}- \mathcal{Q}_b$.}		  & 1.5& 1.1      & 0.03 \footref{dQ} &3.9  & 8.6  \footref{dQ}  \footnote{ For F$_1$-ATPase, we consider a resisting torque ($\tau$ with the unit of pN$\cdot$nm) against the rotation of the motor.}
		\\ \relax  [ATP] ($\mu$M)   			& 210        & 	460 \footref{dQ}    &200& 20   & 17 \footref{dQ}   & 200 & 16 \footref{dQ} 
		\\ $\mathcal{Q}_{\text{min}} ~ (k_B T)$			 &  4.0 		  & 4.5			&9.2 & 6.5	   &  14 & 5.2  & 4.2 
		\\ $\Delta \mathcal{Q}_{\text{min}} ~ (k_B T)$ &   n/a		    & 1.6			&n/a& n/a	& 0   &  2.6 & 0 
		\\ \hline \bottomrule
	\end{tabular}
	
	\label{table_opt}
\end{table*}

\begin{table*}[h!]
	\caption{Parameters determined for the 6-state double-cycle model \cite{Liepelt07PRL}. The unit of rate constants ($\{k_{ij}\}$) is $s^{-1}$ except for $k_{ij}^{bi}$ ($[k_{ij}^{bi}]={\mu \text{M}}^{-1} s^{-1}$).
		The rates in the table are determined for $f=0$.
	}
	\begin{tabular}{| c || c | c | c | c | }
		\midrule
		\hline
		& Kinesin-1 & Kin6AA& KIF17 & KIF3AB \\
		\hline
		$k_{12}^{bi}$ & $2.8$ & 10 & 10 & 10 \\
		$k_{21}$ & $4200$ & 92& 3600 & 500 \\
		$k_{25}$ & $1.6\times 10^6$ & $1.6\times 10^4$  & $4.0 \times 10^4$ & $6.2\times 10^6$ \\
		$k_{52}$& $1.1$ & 3.4 &0.079 & 7.2\\
		$k_{56}$ & $190$& 680& 590& 92\\
		$k_{65}$ & $10$&4.1 & 13& 37\\
		$k_{61}$ & $250$ & 58& 310& 320\\
		$k_{16}$ & $230$ & 260& 1100 & 750\\
		$k_{54}$ & $2.1 \times 10^{-9}$&$4.3 \times 10^{-6}$ & $1.4 \times 10^{-8}$&$6.8 \times 10^{-10}$ \\
		$\theta$ & 	$0.61$ & 0.59 & 0.34 & 0.82 \\
		$\chi_{12}$ & $0.15$ &  0.12& 0.15  & 0.09\\
		$\chi_{56}$ & $0.0015$ & 0.0&0.012 &0.021 \\
		$\chi_{61}$ &$0.11$ &0.18 & 0.17 & 0.16 \\
		\hline
		\bottomrule
	\end{tabular}
	\label{table}
\end{table*}

\begin{table*}[h!]
	\caption{Initial values and constraints applied during the fit of kinesin data using 6-state double-cyclic model. 
		The units are identical to those in Table \ref{table}. 
		For $k_{65}$ and $k_{16}$, we used 7 initial values ($0.001, 0.01, 0.1, 1, 10, 100, 1000$) for the fits. 
	} 
	\begin{tabular}{|| c | l || c| l || c | l || c| l||  }
		\midrule
		\hline
		$k_{12}^{bi}$ & $0.5 \leq 1.8 \leq 10$  & $k_{56}$ & $10 \leq 200 \leq 10^4$ &$k_{61}$ & $10 \leq 200 \leq 10^4$ & $k_{25}$ & $10^4 \leq 3\times 10^5 \leq 10^7$ \\
		$k_{21}$ & $10 \leq 100 \leq 10^4$   & $k_{65}$ & $10^{-4} \leq 10^{[-3,- 2, -1, 0, 1, 2,3]}\leq 10^4$&$k_{16}$ & $0^{-4} \leq 10^{[-3, -2, -1, 0, 1, 2,3]}\leq 10^4$ &  & \\
		\hline
		$\theta$ & 	$0 \leq 0.3 \leq 1$ & $\chi_{12}$ & $0 \leq 0.25 \leq 1$ & $\chi_{56}$ & $0 \leq 0.05 \leq 1$ &$\chi_{61}$ &$0 \leq 0.05 \leq 1$ \\
		\hline
		\bottomrule
	\end{tabular}
	\label{table_fit}
\end{table*}

\begin{table*}[h!]
	\caption{Parameters used for calculation of $\mathcal{Q}$ of myosin-V. The values are obtained from Ref. \cite{Bierbaum:2011:BPJ}. The unit of rate constants ($\{k_{ij}\}$) is $s^{-1}$ except for $k_{ij}^{bi}$ ($[k_{ij}^{bi}]={\mu \text{M}}^{-1} s^{-1}$).
		The rates in the table are determined for $f=0$.
	}
	\begin{tabular}{| c || c | c | }
		\midrule
		\hline
		& Description & value \\
		\hline
		$k_{12}$ & ADP release & 1.2 \\
		$k_{21}^{bi}$ & ADP binding & 4.5  \\
		$k_{23}^{bi}$ & ATP binding & 0.9 \\
		$k_{32}$& ATP release & $2 \times 10^{-5}$ \\
		$k_{34}$ & step & 7000\\
		$k_{43}$ & reverse step &0.65 \\
		$k_{56}^{bi}$ & ATP binding & 0.9 \\
		$k_{65}$ & 	ATP release & $2 \times 10^{-5}$  \\
		$k_{55,f}$ & step (mechanical) &  $1.5 \times 10^{-8}$\\
		$k_{55,b}$ & reverse step (mechanical)  & $1.5 \times 10^{-8}$ \\
		\hline
		\bottomrule
	\end{tabular}
	\label{table_mV}
\end{table*}

\begin{table*}[h!]
	\caption{Parameters used for calculation of $\mathcal{Q}$ of dynein. The values are obtained from Table 3. of Ref. \cite{Sarlah:2014:BPJ}. The unit of rate constants ($\{k_{ij}\}$) is $s^{-1}$ except for $k_{ij}^{bi}$ ($[k_{ij}^{bi}]={\mu \text{M}}^{-1} s^{-1}$).
		The rates in the table are determined for $f=0$.
	}
	\begin{tabular}{| c || c | c | }
		\midrule
		\hline
		& Description & value \\
		\hline
		$k_{12}$ & Pi release & 5000 \\
		$k_{21}^{bi}$ & Pi binding & 0.01  \\
		$k_{23}$ & ADP release & 160 \\
		$k_{32}^{bi}$& ADP binding & 2.7 \\
		$k_{34}^{bi}$ & ATP binding& 2\\
		$k_{43}$ & ATP release &50 \\
		$k_{45}$ & MT release in poststroke state & 500\\
		$k_{54}$ & MT binding in poststroke state & 100\\
		$k_{56}$ & Power stroke & 5000 \\
		$k_{65}$ & 	Reverse stroke & 10  \\
		$k_{67}$ & linker swing to prestroke &  1000\\
		$k_{76}$ & linker swing to poststroke  & 100 \\
		$k_{71}$ & MT binding in prestroke state  &10000  \\
		$k_{17}$ & MT release in prestroke state &500  \\
		\hline
		\bottomrule
	\end{tabular}
	\label{table_dynein}
\end{table*}

\begin{table*}[h!]
	\caption{Polynomial coefficient of $a_i(\tau)=a_i^{(0)} + a_i^{(1)} \tau + a_i^{(2)} \tau^2$ and $b_i(\tau) = b_i^{(0)} + b_i^{(1)} \tau + b_i^{(2)} \tau^2$ used in F$_1$-ATPase $(N=2)$-unicyclic model. The values are obtained from Table 3. of Ref. \cite{Gerritsma:2010}.
	}
	\begin{tabular}{| c || c | c | c| c|}
		\midrule
		\hline
		& $i=k_{12}$ & $i=k_{21'}$ & $i=k_{1'2}$ & Unit \\
		\hline
		$a_i^{(0)}$ & -16.952 & -5.973 & -19.382 & - \\
		$a_i^{(1)}$ & $9.8 \times 10^{-4}$ & $1.7 \times 10^{-4}$ & 0.129 & (pN nm)$^{-1}$\\
		$a_i^{(2)}$ & $5.8 \times 10^{-4}$ & $1.0 \times 10^{-3}$ & $2.8 \times 10^{-4}$& (pN nm)$^{-2}$\\
		$b_i^{(0)}$ & -16.352 & -2.960 & -18.338 &-\\
		$b_i^{(1)}$ & $-6.6 \times 10^{-2}$ & $-2.7 \times 10^{-2}$ & $5.9 \times 10^{-3}$ & (pN nm)$^{-1}$\\
		$b_i^{(2)}$ & $1.0 \times 10^{-3}$ & $3.6 \times 10^{-4}$ & $-2.1 \times 10^{-4}$& (pN nm)$^{-2}$\\
		\hline
		\bottomrule
	\end{tabular}
	\label{table_f1}
\end{table*}
\begin{table*}[h!]
	\caption{Parameters of (N=4)-unicyclic model of kinesin-1. The values are obtained from our previous study \cite{Hwang2017JPCL}. The unit of $\{u_n\}$ and $\{w_n\}$ is $s^{-1}$ except for $u_1^o$ ($[u_1^o]={\mu M}^{-1} s^{-1}$).}
	\begin{tabular}{ |c | l || c| l || c | l || c| l |  }
		\hline
		$u_1^0$ & $2.3$ & $u_2$ & $600$ & $u_3$ & $400$&$u_4$ & $190$ \\
		$\theta_{1}^{+}$ & $0.00$ & $\theta_{2}^{+}$ & $0.04$ & $\theta_{3}^{+}$ & $0.01$ &$\theta_{4}^{+}$ &$0.02$ \\
		$w_1$ & $20$  & $w_2$ & $1.4$ & $w_3$ & $1.7$ & $w_4$ & $120$ \\
		$\theta_{1}^{-}$ & $0.14$ &  $\theta_{2}^{-}$ & $0.15$& $\theta_{3}^{-}$ & $0.5$ &$\theta_{4}^{-}$ & $0.14$\\
		\hline
	\end{tabular}
	\label{table_kinesin_uni}
\end{table*}

\begin{table*}[h!]
	\caption{Parameters for ($N=2$)-unicyclic model of myosin-V. The values are obtained from Eqs. (12), (13) in Ref. \cite{Kolomeisky:2003:BPJ}.
		The unit of rate constants ($\{u_i, w_i\}$) is $s^{-1}$ except for $k$ and $k'$ ($[k, k']={\mu \text{M}}^{-1} s^{-1}$).
		The rates in the table are determined for $f=0$.
	}
	\begin{tabular}{| c || c | }
		\midrule
		\hline
		$k$ & 0.70 \\
		$w_1$ & $6 \times 10^-6$  \\
		$u_2$ & 12  \\
		$k'$& $5.0 \times 10^{-6}$ \\
		$\theta_1^+$ & -0.01\\
		$\theta_1^-$ & 0.045 \\
		$\theta_2^+$ & 0.385 \\
		$\theta_2^-$ & 0.58\\
		\hline
		\bottomrule
	\end{tabular}
	\label{table_mV_uni}
\end{table*}

\end{document}